\documentclass[referee]{aa}
\usepackage{ae,german}
\usepackage{pstricks}
\usepackage{color}
\usepackage{pdfpages}
\usepackage{textcomp}
\usepackage{graphicx, graphics, epsfig, psfrag}
\usepackage{wasysym}
\usepackage{amsfonts, amsmath}
\usepackage[english]{babel}
\usepackage{tikz}
\usepackage{natbib}
\usepackage{slashed}
\usepackage{ads}
\usepackage[normalem]{ulem}

\newenvironment{ita}{\itshape}{\upshape}
\newenvironment{bld}{\bfseries}{\mdseries}

\newenvironment{grt}{\greektext}{\latintext}

\newcommand{\bit}{\begin{ita}}
\newcommand{\eit}{\end{ita}}
\newcommand{\bbl}{\begin{bld}}
\newcommand{\ebl}{\end{bld}}
\newcommand{\bgr}{\begin{grt}}
\newcommand{\egr}{\end{grt}}
\newcommand{\bean}{\begin{align}}
\newcommand{\eean}{\end{align}}
\newcommand{\bas}{\begin{align*}}
\newcommand{\eas}{\end{align*}}
\newcommand{\beq}{\begin{equation}}
\newcommand{\eeq}{\end{equation}}
\newcommand{\beqs}{\begin{equation*}}
\newcommand{\eeqs}{\end{equation*}}

\newcommand{\ben}{\begin{eqnarray}}
\newcommand{\een}{\end{eqnarray}}
\newcommand{\bens}{\begin{eqnarray*}}
\newcommand{\eens}{\end{eqnarray*}}

\newcommand{\AU}{\text{AU}}
\newcommand{\meter}{\text{m}}

\newcommand{\second}{\text{s}}
\newcommand{\Me}{\text{M}_{\oplus}}
\newcommand{\Ms}{\text{M}_{\odot}}
\newcommand{\MJ}{\text{M}_{Jup}}

\newcommand{\Myr}{\text{Myr}}
\newcommand{\Gyr}{\text{Gyr}}
\newcommand{\yr}{\text{yr}}
\newcommand{\lsim}{\apprle}
\newcommand{\gsim}{\apprge}

\newcommand{\lp}{\left(}
\newcommand{\rp}{\right)}
\newcommand{\lb}{\left[}
\newcommand{\rb}{\right]}

\newcommand{\mc}[1]{\mathcal{#1}}

\newcommand{\symba}{\textit{SyMBA}}
\newcommand{\bal}{\begin{align}}
\newcommand{\eal}{\end{align}}
\newcommand{\bals}{\begin{align*}}
\newcommand{\eals}{\end{align*}}

\renewcommand{\deg}{\!\phantom{.}^{\circ}}
\newcommand{\rv}{K\gsim10\meter\second^{-1}}
\newcommand{\fwat}{f_{{H_2O,core}}}

\begin{document}

\title{Theoretical models of planetary system formation}
\subtitle{II. Post-formation evolution}

\author{S.Pfyffer\inst{1} \and Y.Alibert\inst{1,2} \and W.Benz\inst{1} \and D.Swoboda\inst{1}}

\institute{Physikalisches Institut \& Center for Space and Habitability, Universit\"at Bern, Sidlerstrasse 5, CH-3012 Bern, Switzerland\\ \email{samuel.pfyffer@space.unibe.ch, WBenz@space.unibe.ch, yann.alibert@space.unibe.ch, david.swoboda@space.unibe.ch} \and
Observatoire de Besançon, 41 Avenue de l'Observatoire, F-25000 Besançon, France\\ \email{alibert@obs-besancon.fr}
}

\date{Received xyz; Accepted uvw}

\abstract{}{We extend the results of planetary formation synthesis by computing the long-term evolution of synthetic systems from the clearing of the gas disk into the dynamical evolution phase.}{
We use the symplectic integrator \symba\ (\cite{Duncan1998}) to numerically integrate the orbits of planets for $100\Myr$, using the populations in \cite{Alibert2013} as initial conditions.}{We show that within the populations studied, mass and semi-major axis distributions experience only minor changes from post-formation evolution. We also show that, depending upon their initial distribution, planetary eccentricities can statistically increase or decrease as a result of gravitational interactions. We find that planetary masses and orbital spacings provided by planet formation models do not result in eccentricity distributions comparable to observed exoplanet eccentricities, requiring other phenomena such as e.g. stellar fly-bys to account for observed eccentricities.}{}

\maketitle

\section{Introduction}

Since the pioneering discovery of the first exoplanet orbiting a main-sequence star by \cite{51Pegb1995}, the number of discovered exoplanets has increased to over 1000 planets with a plethora of characteristics challenging planet formation models. The latest generation of core accretion models such as the planet formation model in \cite{Alibert2013} (see  \cite{Alibert2005}, \cite{Mordasini2009,Mordasini2009b}, \cite{Fortier2013} and \cite{Alibert2013}, henceforth referred to as A05, M09, M09b, F13 and A13, respectively) or the semi-analytical method of Ida and Lin (see \cite{Ida2004,Ida2010,Ida2013}) permits the study of planet formation not only for isolated planets, but for planets forming concurrently in the same system. This allows for the study of the effects of mutual perturbation between growing protoplanets. Thus, planetary characteristics such as eccentricity and architectural features such as period ratios - both depending on mutual interactions - can be compared against the distributions for observed exoplanets.\\
While the distributions of semi-major axes and planetary masses are generally well reproduced by planet formation models (see e.g. M09 and M09b), the eccentricities of synthetic giant planets are statistically smaller than those for observed exoplanets (see e.g. A13). This has been attributed to the damping effect of the gas disk during the formation phase. Studies by e.g. \cite{Juric2008} showed that observed eccentricities can be reproduced by following the gravitational post-formation evolution of systems, but the simulations relied for their initial conditions on ad-hoc architectures of the initial systems. It is not clear whether such architectures can actually result from the formation process. It is therefore uncertain whether dynamical instabilities suppressed while the gas disk is still present can account for the observed exoplanet eccentricities or if other effects such as e.g. perturbations from stellar fly-bys (see e.g. \cite{Malmberg2009}) are necessary as a source of eccentric planets. \cite{Thommes2008}, using a code coupling a 1D gas disk evolution code with \symba, found that for some disk parameters, giant planets with substantial eccentricities can be formed by planet-planet and planet-disk interactions. Such planets are also obtained by A13, but their occurrence rates in the planet populations are relatively low and thus insufficient to match observed exoplanet distributions. \\
N-body simulations of the orbital evolution of (proto-)planets in gas disks (see e.g. \cite{McNeil2005,Lee2009,Matsumura2010}) found that a large number of planets ends up locked in low-order mean-motion resonances due to resonant migration. The large sample of exoplanets observed by the Kepler spacecraft, however, does not show a similar enrichment in commensurable period ratios. Therefore, in formation models including both N-body dynamics and planet-disk interactions, a mechanism to break out of resonance is necessary to reduce the fraction of commensurable orbits. \\
In the present paper, we study the impact of post-formation dynamical evolution on planetary systems formed with the model of A13. We investigate whether the post-formation evolution alone can account for the observed statistical increase in eccentricities. Previous studies presented in M09b and A13 followed the formation of planets until the dispersal of the protoplanetary gas disk and typically lasted for up to $T_{\text{disk}}\sim10\Myr$. We improve on these results by following the subsequent gravitational evolution of synthetic planetary systems for an additional $100\Myr$ after the dispersal of the protoplanetary disk, partially bridging the time gap between the formation era and the actual observation epoch. Between the time of disk dispersal and the long-term dynamically stable configurations of mature planetary systems, dynamical instabilities hitherto suppressed by the gas disk through eccentricity and inclination damping can arise, resulting in scatterings, collisions and ejections of planets from their respective systems.\\
In Section \ref{sec:plfm}, we will briefly review the planet formation model used in this paper and discuss the procedure we employed to compute the post-formation evolution and the planetary populations studied in Section \ref{sec:popsyn}. We will discuss the effect on the evolved eccentricity distributions of a number of initial condition parameters as well as small stochastic perturbations in Section \ref{sec:ecc}, while the post-formation evolution effects in general are presented in Section \ref{sec:ltev}. Finally, we provide a brief discussion and conclusion.

\section{Planet formation model}\label{sec:plfm}

In the present work, we study the post-formation evolution decoupled from the planet formation process, numerically integrating the orbits of planets generated with a planet formation model. We use the planet formation model described in F13 and A13, based on earlier works (see A05 and M09), to generate populations of planetary systems in a Monte Carlo approach from initial parameters calibrated to observations (given in A13). The Bernese model self-consistently simulates the evolution of the gas and planetesimal disks, computes the solid and gas accretion together with the internal structure of planets, and accounts for both the gravitational interaction between planets and the interaction of planets with the surrounding gas disk.\\ The disk structure and evolution is modelled using a 1+1D-model computing the radial and vertical structure of the disk. The time evolution of the disk is obtained from the diffusion equation for the gas disk surface density. A detailed description of the disk structure and evolution model is given in \cite{Papaloizou1999} and A05. The interactions between growing protoplanets and the gas disk are computed using the prescriptions of \cite{Fogg2007} for Type I migration in the case of an isothermal disk and by setting the eccentricity damping proportional to the semi-major axis damping for planets in Type II migration. For non-isothermal disks, the migration and damping prescriptions of \cite{Paardekooper2010,Paardekooper2011} and \cite{Dittkrist2014} are used. A more detailed description of planet-disk interactions is given in A13.

\subsection{Evolution procedure}\label{sec:ltevp}
To follow the long-term evolution of the systems generated with our population synthesis code, we use the symplectic integrator \symba\ (\cite{Duncan1998}). The post-formation evolution of multiplanetary population syntheses is then obtained in the following way:
\begin{enumerate}
\item A population of planetary systems is generated using the planet formation model from A13. In this model, planets are followed dynamically until the gas disk mass decreases to $M_{\text{disk}}\leq 10^{-5}\Ms$. 

\item The system is then evolved over $100\Myr$ using the symplectic integrator \symba.
\end{enumerate}
The chosen timescale for evolution of $100\Myr$ is a compromise between the time required to explicitly integrate a system in time and the number of systems we need to integrate to obtain a statistically meaningful result. The chosen integration time of $100\Myr$ is sufficiently large to allow for dynamical effects (cf. \cite{Juric2008}, where they find a rapid drop in dynamical activity for $T>100\Myr$ in integrations running up to $\sim 1\Gyr$) acting on timescales longer than the disk lifetime such as e.g. for secular resonances, orbit crossing and collisions, where the timescales are typically of the order of a few $\Myr$.\\
To reduce computational load and to focus on potentially observable systems, we 
set a 
minimum mass of the most massive planet ($m_{\text{start}}$) below which the system is not integrated, i.e. if no planet with $m_p\geq m_{\text{start}}$ is present in the system, the system is omitted from long-term integration. For the simulations studied in this paper, we set $m_{\text{start}}$ to $1\Me$. This choice is also motivated by results from A13, where a convergence in the number of planets with a given mass is shown for planets with $M_p\geq5\Me$ as a function of the initial number of embryos and initial conditions. For planets with $M_p\leq5\Me$, significant effects from initial conditions on their occurrence rates and properties cannot be excluded. These uncertainties increase with further decreasing mass.\\
Collisions during the post-formation evolution are treated as perfect mergers, occurring in  \symba\ if the separation $d_{ij}$ between two objects is less than the sum of their physical radii. Note that we do not distinguish between core and envelope radius in our treatment of collisions.

\section{Population synthesis}\label{sec:popsyn}

To test the effects of different parameters in the formation model, we generated several populations of planetary systems with the planet formation code (see Sec. \ref{sec:plfm}), using the model described in A13. 
\\
A summary of the populations we studied is given in Table \ref{tab:popov_all}. Below, each population is briefly described. We use a criterion based on the radial velocity semi-amplitude $K$ to select the planets in each population which we compare to a similarly selected sample of observed exoplanets, with $\rv$. We do not, however, apply a cutoff based on orbital periods as the fraction of planets with orbital periods larger than $\sim20\yr$ - roughly the time since measurements of $K\leq10\meter\second^{-1}$ would have been available from e.g. the ELODIE spectrograph -  is less than $\sim3\%$ except for the population ND (see below), where the fraction is of $\sim13\%$.
\begin{table*}[htbp]\caption{Overview of population syntheses. $n_e$ denotes the number of embryo seeds, $n_{sys}$ the number of systems, $n_p$ (resp. $n_p$ @$10\meter\second^{-1}$) the number of  planets (resp. with a radial-velocity semi-amplitude $K\geq10\meter\second^{-1}$) and $n_{coll}$ and $n_{eje}$ the number of planets removed due to collisions and ejections. The population designations refer to the reference population with 10 embryos and full damping (R), the population with 5 and 20 embryos (E5, E20), the population with no (ND) and reduced eccentricity and inclination damping (RD), and the population with a stochastic perturbation to the semi-major axes (RA) or to the inclinations (RI).}\label{tab:popov_all}
\centering
\begin{tabular}{|c|c|c|c|c|c|c|c|} \hline
 & R & E20 & E5 & ND & RD & RA & RI \\ \hline
initial $n_e$ & 10 & 20 & 5 & 10 & 10 & 10 & 10 \\ \hline
$n_{sys}$ & 344 & 228 & 483 & 264 & 343 & 344 & 413 \\ \hline
initial $n_p$ & 2834 & 3241 & 2171 & 1686 & 2480 & 2834 & 3309 \\ \hline
final $n_p$ & 2519 & 2238 & 2129 & 1236 & 2155 & 2389 & 2921 \\ \hline
initial $n_{p}$ @10$\meter\second^{-1}$ & 206 & 160 & 241 & 143 & 217 & 206 & 253 \\ \hline
final $n_{p}$ @10$\meter\second^{-1}$ & 204 & 153 & 237 & 132 & 191 & 202 & 244 \\ \hline
$n_{coll}$ & 282 & 902 & 000 & 318 & 225 & 378 & 329 \\ \hline
$n_{eje}$ & 22 & 84 & 000 & 112 & 86 & 48 & 45 \\ \hline
\end{tabular}

\end{table*}

\subsection{The reference population}\label{sec:pop_R}

The reference population against which we compare the results of changes to the initial conditions of the planet formation process (populations ND, RD, E, see below) and stochastic perturbations (populations RA and RI, see below) was generated using $10$ initial embryo seeds and disk masses, disk lifetimes and dust-to-gas ratios drawn randomly from the parameter distributions described in A13. The disk masses vary from $M_{d}=0.001\Ms$ to $M_{d}=0.3\Ms$, with a dust-to-gas ratio scaled to the metallicity of the star. The gas disk masses are derived from \cite{Andrews2010}, where a dust-to-gas ratio of $f_{d/g}=0.01$ is assumed. The solid disk mass is computed by scaling the mass of the solid disk according to the randomly drawn stellar metallicity from a list of $\sim1000$ CORALIE targets (see also  A13) with a new $f_{d/g}$. In our simulations, most disks are low-mass, with $\sim62\%$ of all disks having $M_d\leq0.03\Ms$. Disk lifetimes $T_{\text{disk}}$ are chosen from an exponentially decaying cumulative distribution function with characteristic time $T=2.5\Myr$ (see \cite{Mamajek2009}). The photoevaporation rate is then adjusted to yield $M_{d}=10^{-5}\Ms$ at $T=T_{\text{disk}}$. Finally, embryos are seeded throughout the disk. All embryos have a starting mass of $M=10^{-2}\Me$ and an initial semi-major axis randomly drawn from a log-uniform distribution between $0.1\AU$ and $20\AU$, with an added constraint that embryos are located at least 10 times their mutual Hill radius from each other. After all systems in the population have reached $M_{\text{disk}}\leq10^{-5}\Ms$, the systems containing no planet with $M_p\geq m_{\text{start}}$ were filtered out, while the remaining systems were integrated for an additional $100\Myr$ to follow the post-formation evolution under the mutual gravitational interactions of the planets alone (i.e. any residual gas or planetesimal disk and tidal interactions were not included). This resulted in a population of 2834 planets in 344 systems being evolved beyond the formation phase, with 2519 planets remaining after $100\Myr$. 

\subsection{Modified damping timescales - populations ND and RD}\label{sec:pop_NDRD}

As there are some uncertainties about the timescales of inclination and eccentricity damping of planets due to the interactions with protoplanetary disks (see e.g. \cite{Bitsch2010}), we performed population syntheses with the same initial conditions as in the reference case, but modifying the damping timescales. The resulting populations without any eccentricity or inclination damping (population ND) and with the damping timescales increased by a factor of $10$ with respect to the damping timescale prescription given in \cite{Alibert2013} (population RD) respectively resulted in a population of 1853 planets in 264 systems, with 1492 planets remaining after $100\Myr$ (population ND) and in a population of 2480 planets in 343 systems, of which 2155 planets remained after $100\Myr$ (population RD).

\subsection{Different initial number of embryo seeds}\label{sec:pop_E}

The number of embryos is a free parameter of the model, and \cite{Alibert2013} have already discussed the effects of changing the number of initial embryo seeds on the distributions of planetary masses and semi-major axes as well as on period ratios of neighboring planet pairs for the formation phase until the disk has dispersed. However, as the number of embryos increases, we also expect an increase in dynamical interactions. We therefore also computed the post-formation evolution of a population with the same initial conditions as the reference case, but with the number of initial embryo seeds increased to 20 (population E20), as well as a population with the same initial conditions as the reference population, but with only 5 initial embryo seeds (population E5). To maintain a similar total number of planets, a lower number of systems was simulated, resulting in a population of 3241 planets in 228 systems, of which 2238 planets remain after $100\Myr$, whereas a slightly larger number of systems was simulated for the population E5, resulting in 2171 planets in 483 systems, of which 2129 planets remain after $100\Myr$.

\subsection{Additional stochastic perturbations}\label{sec:pop_S}

In the Bernese planet formation model, the migration prescriptions assume that the disk is azimuthally symmetric and homogeneous. Moreover, changes due to e.g. turbulence (see e.g. \cite{Johnson2006}) which can affect the migration of low- to intermediate-mass planets are not included in the model. To simulate the effect of stochastic perturbations on the planets, we computed the post-formation evolution of the reference population, but added a random perturbation within $\pm$1\% to the semi-major axis of each planet (population RA) before computing the post-formation evolution. This is not intended as a self-consistent scheme as the perturbation bears no effect on the formation process, but rather as cursory investigation into the effect of semi-major axis perturbation on the subsequent evolution of planetary systems. However, we also computed a second population with stochastic perturbations (the population RI), but in a self-consistent way, by adding a perturbation to planetary inclinations during the formation phase. The inclinations are perturbed by drawing a random inclination from a uniform distribution in $\lb0\deg,1\deg\rb$ for planets with $i\leq 1\deg$.

\section{Results: effect of initial conditions on the eccentricity distribution}\label{sec:ecc}

We used the post-formation evolution of a sample of 344 systems with 10 initial embryos each as the baseline against which we compared the effect of altering different aspects of the planet formation model, namely the number of embryos and the efficiency of the eccentricity (and inclination) damping due to interactions with the surrounding gas disk.

\subsection{The initial number of embryos}\label{sec:ene}

\begin{figure}
\includegraphics[width=0.5\textwidth]{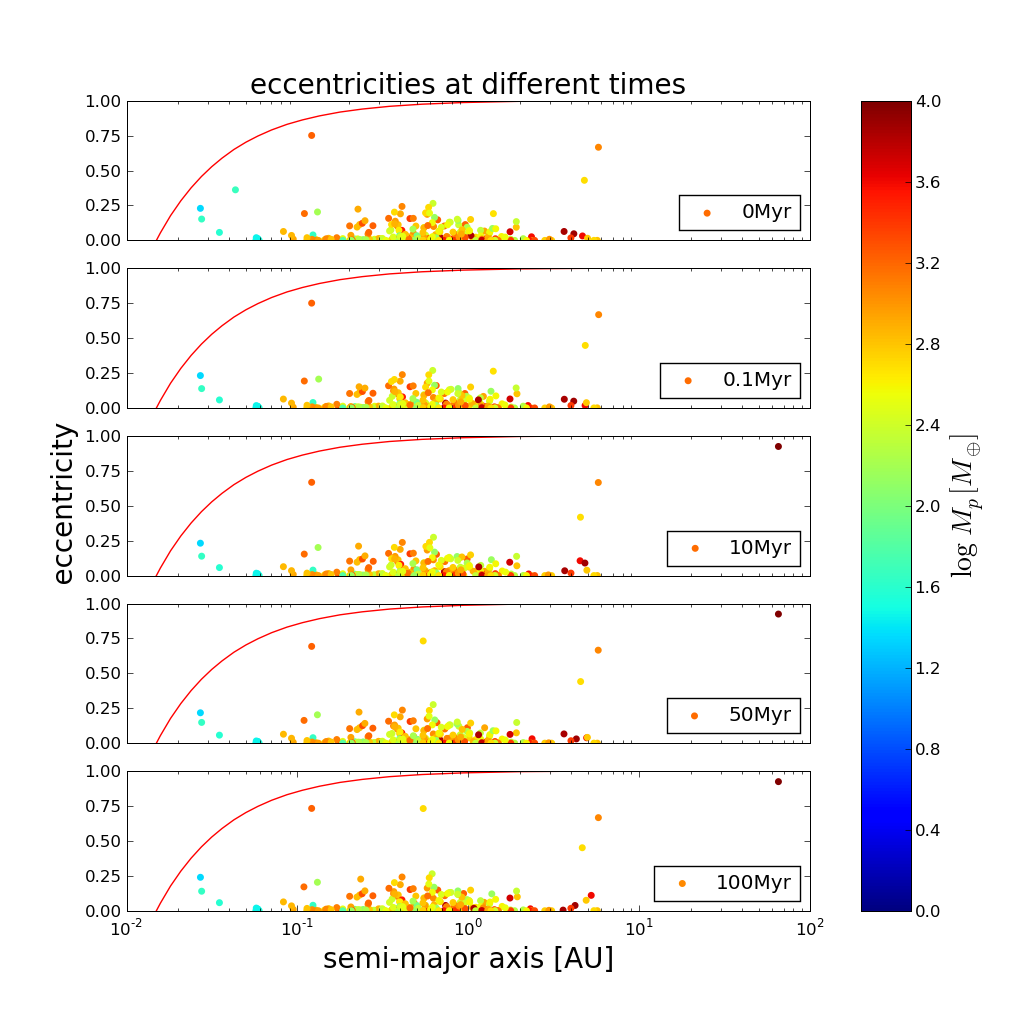}
\caption{Semi-major axis versus eccentricity for all planets in the 5 embryo population (E5) which would produce a radial velocity semi-amplitude of $K\gsim10\meter\second^{-1}$. The upper panel corresponds to the initial distribution, the lower panel to the distribution after $100\Myr$. The colorscale corresponds to the logarithm of the planetary mass in Earth masses, while the red solid line denotes the eccentricity at a given semi-major axis for which the periastron separation results in the planet's removal from the simulation.}\label{fig:CD2159_ae_both}
\end{figure}

\begin{figure}
\includegraphics[width=0.5\textwidth]{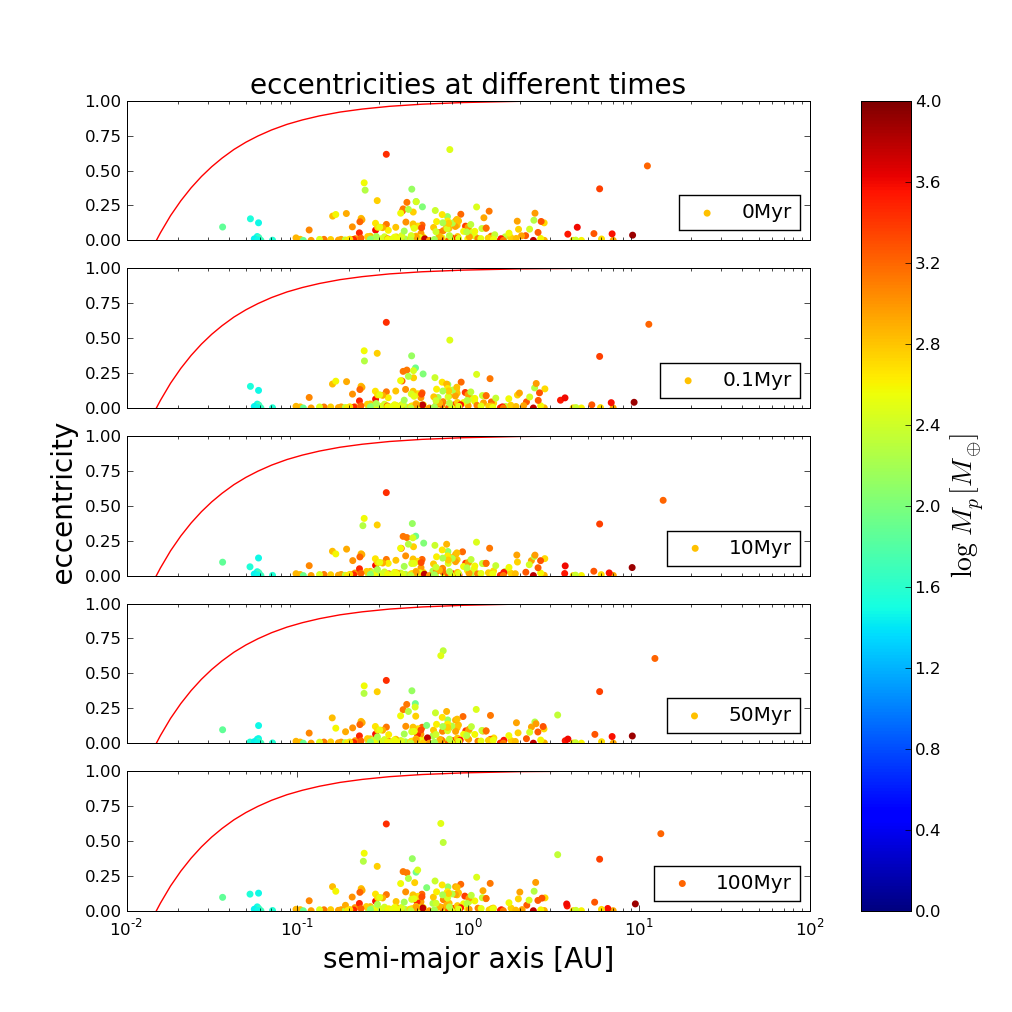}
\caption{Same as Fig. \ref{fig:CD2159_ae_both} for the reference population (R).}\label{fig:CD2133_ae_both}
\end{figure}

\begin{figure}
\includegraphics[width=0.5\textwidth]{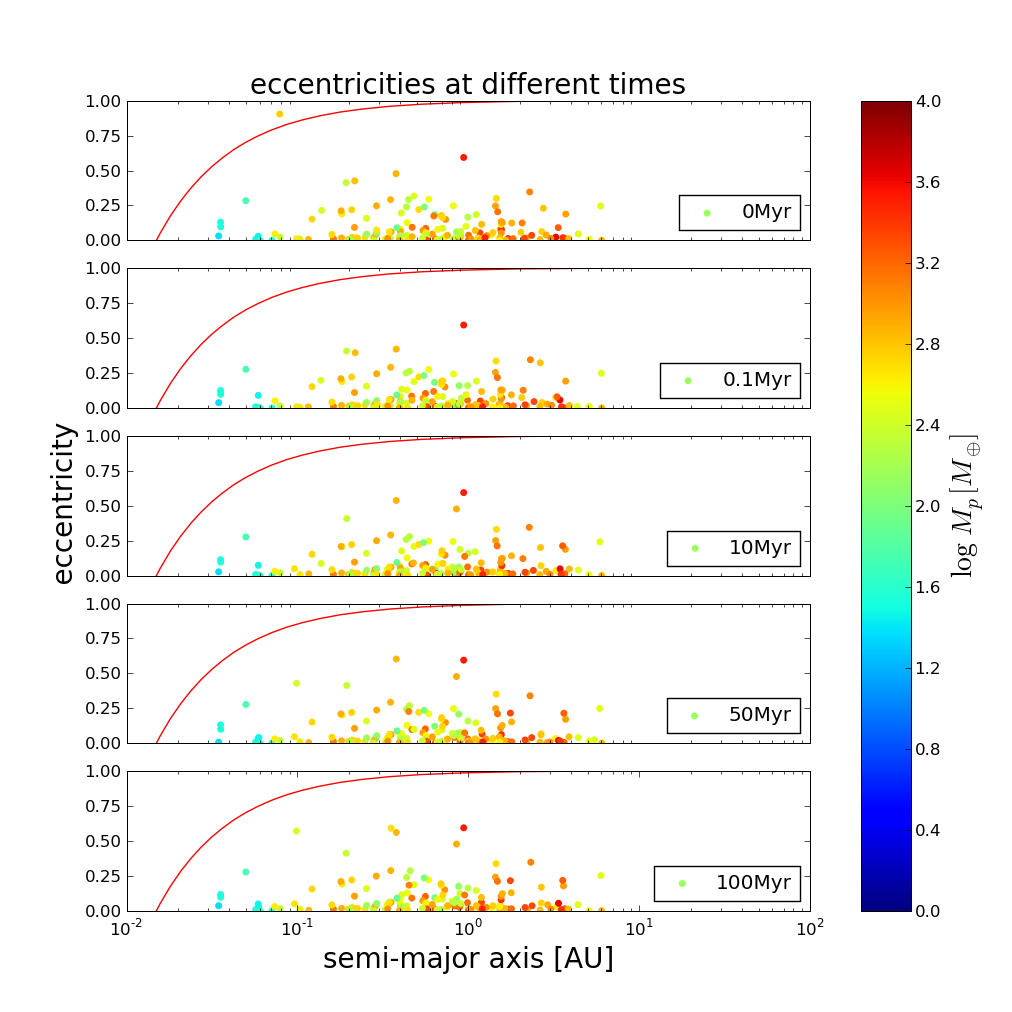}
\caption{Same as Fig. \ref{fig:CD2159_ae_both} for the population with 20 initial embryos (E20).}\label{fig:CD2160_ae_both}
\end{figure}

In A13, the impact of the initial number of embryo seeds on the distributions of mass and semi-major axis was studied. The distributions for 5, 10 and 20 initial embryo seeds show a converging trend both with respect to the distribution of planetary masses (for planets with $M_p\gsim 10\Me$) and the distribution of semi-major axes. In the present work, we compared the distribution of eccentricities between the different populations, both after the initial formation phase which typically lasts a few $\Myr$ until the gas disk is dissipated, and after a subsequent gravitational evolution lasting for $100\Myr$ after the time of disk dissipation. We find that unlike for the mass and semi-major axes, the eccentricities and their evolution differ between the populations with 5 (Figure \ref{fig:CD2159_ae_both}), 10 (Figure \ref{fig:CD2133_ae_both}) and 20 (Figure \ref{fig:CD2160_ae_both}) initial embryo seeds. The median eccentricity at the time of disk dissipation increases with increasing number of embryos from $e=0.019$ (E5) to $e=0.030$ (R) to $e=0.039$ (E20) for planets with $K\geq10\meter\second^{-1}$. The full populations including the lower-mass planets follows a similar trend, albeit at much lower eccentricities, with $e=4.1\cdot10^{-4}$, $e=5.8\cdot10^{-4}$ and $e=8\cdot10^{-4}$. The different populations are also affected differently by the post-formation evolution. While all three populations show an increase in eccentricity for the full populations (median eccentricities of $e=4.6\cdot10^{-4}$, $e=1.2\cdot10^{-3}$ and $e=5.8\cdot10^{-3}$, respectively), the populations of planets with $K\geq10\meter\second^{-1}$ show a slight decrease in median eccentricity for the population with 5 ($e=0.017$) and 10 embryos ($e=0.029$), whereas the population with 20 embryos shows a slight increase in median eccentricity ($e=0.042$).\\
We note that in all three populations the median eccentricities (and maximum eccentricities) are larger for the populations restricted to planets with $\rv$. On one hand, the eccentricity damping in our model is proportional to the damping in semi-major axis and thus most efficient for planets with $m_p\lsim1-10\Me$, the mass around which the planets enter runaway gas accretion and ultimately transition into the slower Type II migration. By consequence, the more massive planets with $\rv$ migrating in Type II are less strongly damped than the lower-mass planets.
On the other hand, the eccentricities of our planet populations are due to dynamical interactions, i.e. resonant excitations and secular perturbations as well as close encounters. Previous numerical and analytical studies (see e.g. \cite{Ida2010} and \cite{Ida2013} and references therein) have found that planetary eccentricities from two-body encounters correspond to a Rayleigh distribution multiplied with the mass-weighted escape eccentricity of the encounter, the latter being defined as $e_{\text{esc},ij}=\frac{v_{esc,ij}}{v_{K}}=\sqrt{\frac{\lp m_i+m_j\rp a}{M_{\star}\lp R_i+R_j\rp}}$, where $a=\sqrt{a_i a_j}$ is the geometric mean semi-major axis of the planets $i$ and $j$, $m_i$, $m_j$ and $M_{\star}$ respectively denote the mass of both planets and the star and where $R_i$ and $R_j$ are the planetary radii. The prescription for the resulting maximum post-encounter eccentricity is then given by $e_{\text{max},j}=\frac{m_i}{m_i+m_j}\mc R\: e_{\text{esc},ij}$, where $\mc R$ denotes the random variate from the Rayleigh distribution. For an encounter between low-mass planets, the escape eccentricity typically remains small, such that encounters between low-mass planets only lead to small increases in eccentricity. In contrast, encounters between a low-mass planet ($j$) and a massive planet ($i$) can rapidly result in $e_{\text{max},j}>1$ and thus in the ejection of the low-mass planet without substantially increasing the eccentricity of the massive planet. Encounters between two massive planets, while still potentially resulting in ejections, increase the eccentricities of both planets and thus produce a population of massive planets with some eccentricity. This bias against eccentric, low-mass planets is further compounded in our simulations by a large number of low-mass planets in systems without any massive companion, such that low-mass planets predominately interact with other low-mass planets. The correlation between planetary masses and eccentricities, however, is also found in observed exoplanets (see e.g. \cite{Ida2013}).


\begin{figure}
\includegraphics[width=0.5\textwidth]{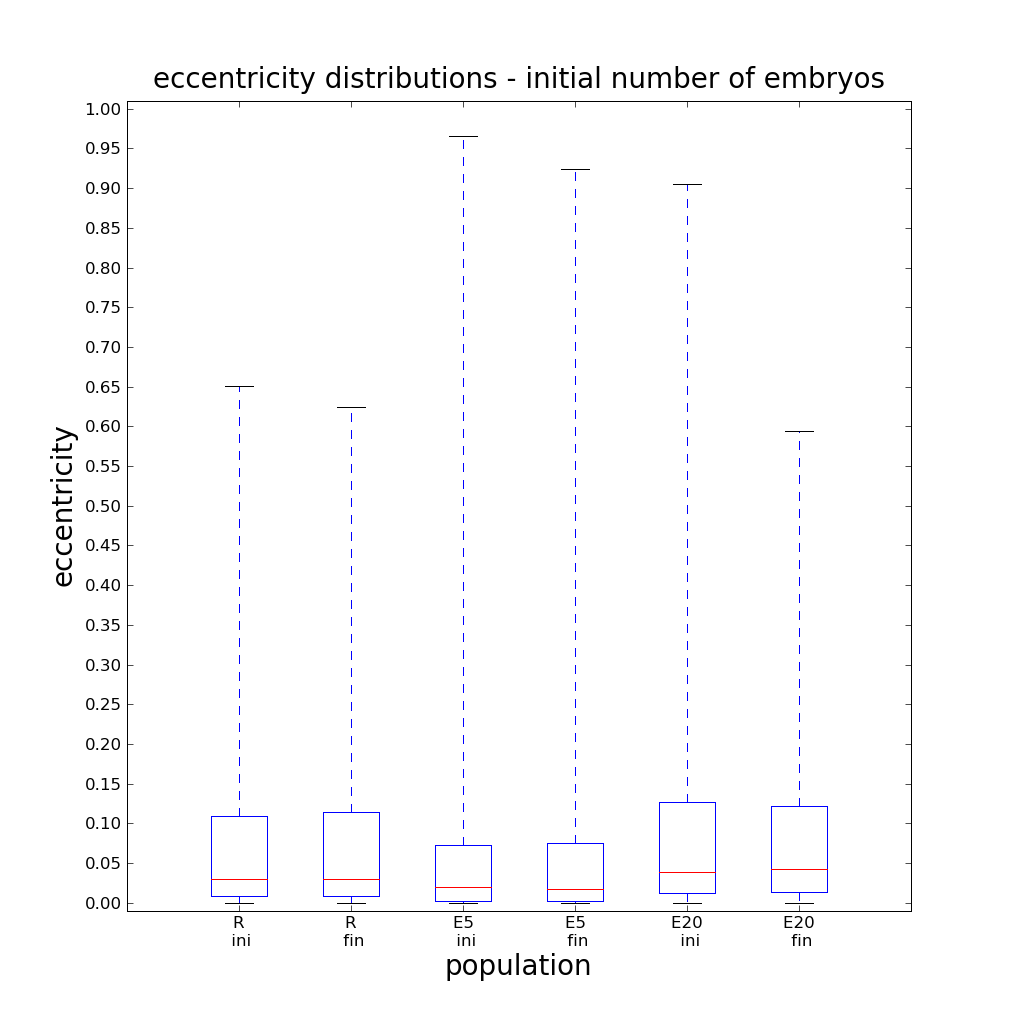}
\caption{Comparison of the eccentricity distributions of the populations R, E5 and E20, limited to planets with $K\gsim10\meter\second^{-1}$. The red lines correspond to the median values, the blue boxes respectively denote the interquartile ranges while the whiskers correspond to the minimum and maximum values.}\label{fig:box_CD_ex_10ms}
\end{figure}

\begin{figure}
\includegraphics[width=0.5\textwidth]{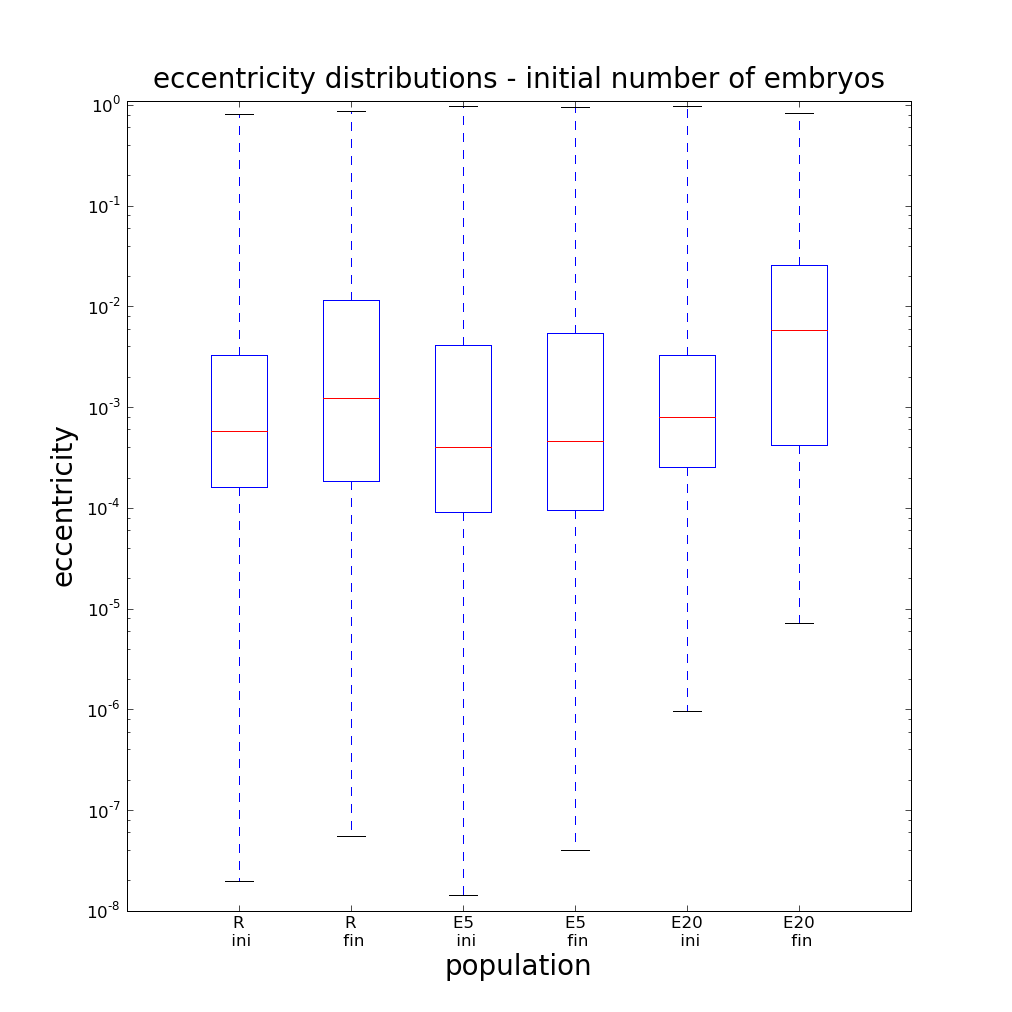}
\caption{Comparison of the eccentricity distributions of the full populations R, E5 and E20. The red lines correspond to the median values, the blue boxes respectively denote the interquartile ranges while the whiskers correspond to the minimum and maximum values.}\label{fig:box_CD_ex_all}
\end{figure}

\subsection{Eccentricity damping}

A13 presented results for population syntheses with modified eccentricity damping timescales and found that including no eccentricity and inclination damping (ND) or increased damping timescales (RD) quite naturally led to populations with significantly larger eccentricities. We computed the post-formation evolution of the same populations to study the effect of reduced eccentricity damping during formation on the final architecture of planetary systems. We found that in both cases with modified damping, the median eccentricities of the populations of planets with $K\gsim10\meter\second^{-1}$ decrease over time, from $e=0.148$ to $e=0.129$ (ND) and from $e=0.102$ to $e=0.095$. For the full populations, the median eccentricity decreases in the case of no damping, dropping to $e=0.066$ from $e=0.122$ at the time of disk dispersal, whereas the median eccentricity increases from $e=1.0\cdot10^{-3}$ at the time of disk dispersal to $e=2.2\cdot10^{-3}$ after $100\Myr$.\\
Contrary to our original expectation that eccentricities would grow throughout the post-formation evolution, we find that in both populations, the median eccentricity for planets with $\rv$ decreases over $100\Myr$. 
This is explained by the fact that starting with larger post-formation eccentricities and inclinations, the systems are more dynamically active, which leads to more collisions and ejections. The latter typically only affect planets with eccentricities close to or larger than unity, as we consider a planet to be ejected from the system if its heliocentric distance $r_p\geq1000\AU$ and only very few planets are scattered onto wide orbits with moderate eccentricities. This eventually leads to a decrease in eccentricity not fully compensated by enventual excitation of the remaining body in a two-body encounter (see Sec. \ref{sec:ene}). In addition to the loss of eccentric planets through ejections, collisions also reduce the eccentricities. This can be more readily understood from evaluating the angular momentum deficit of a planet, introduced by \cite{Laskar1997} and defined as
\beq
\mc C_i=m_i\sqrt{GM_{\star}a_i}\lp1-\sqrt{1-e_i^2}\cos{i_i}\rp,
\eeq
with semi-major axis $a_i$, eccentricity $e_i$ and inclination $i_i$ and with $m_i$ and $M_{\star}$ respectively denoting the planet and host star mass. The AMD characterizes the degree of non-linearity in a given orbit. For purely secular systems, the total angular momentum deficit, given by the sum over all individual $\mc  C_i$, is conserved. \cite{Laskar2000} notes that in a fully elastic collision, the total angular momentum deficit of the colliding pair decreases. By consequence, the inclination and/or eccentricity of the resulting planet tend to be lower than the eccentricities of the parent bodies. As the planets in our simulations are generally on well-separated orbits, only eccentric planets undergo collisions, thus again removing the more eccentric planets from the population. \\
The relatively large fraction of collisions is due to large initial populations of planets with Safronov numbers $\Theta<1$, for which the preferred outcome of close encounters are collisions rather than ejections. These large fractions ($>94\%$ for both ND and RD) are due to a large number of both close-in and low-mass planets for which the local orbital escape velocity is larger than the surface escape velocity, such that a scattering by the planet is not likely to eject the scattered planet from the system.

\begin{figure}
\includegraphics[width=0.5\textwidth]{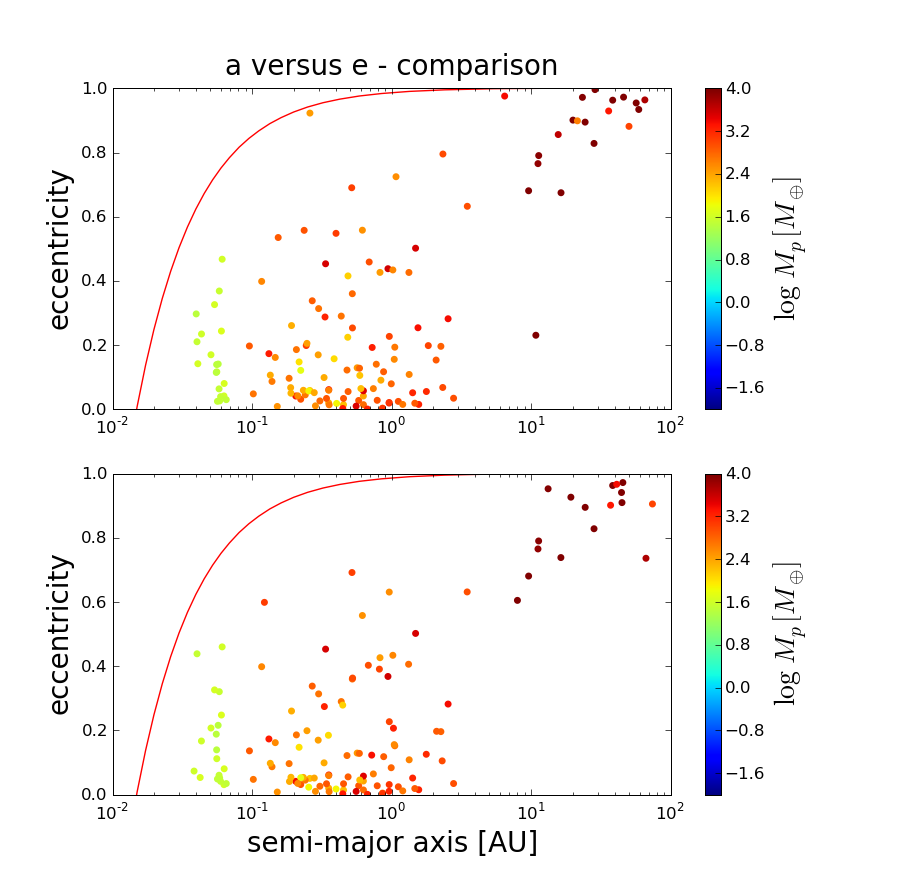}
\caption{Semi-major axis versus eccentricity for all planets in the 10 embryo population without eccentricity damping which would produce a radial velocity semi-amplitude of $K\gsim10\meter\second^{-1}$. The upper panel corresponds to the initial distribution, the lower panel to the distribution after $100\Myr$. The colorscale corresponds to the logarithm of the planetary mass in Earth masses, while the red solid line denotes the eccentricity at a given semi-major axis for which the periastron separation results in the planet's removal from the simulation.}\label{fig:CD2156_ae_both}
\end{figure}

\begin{figure}
\includegraphics[width=0.5\textwidth]{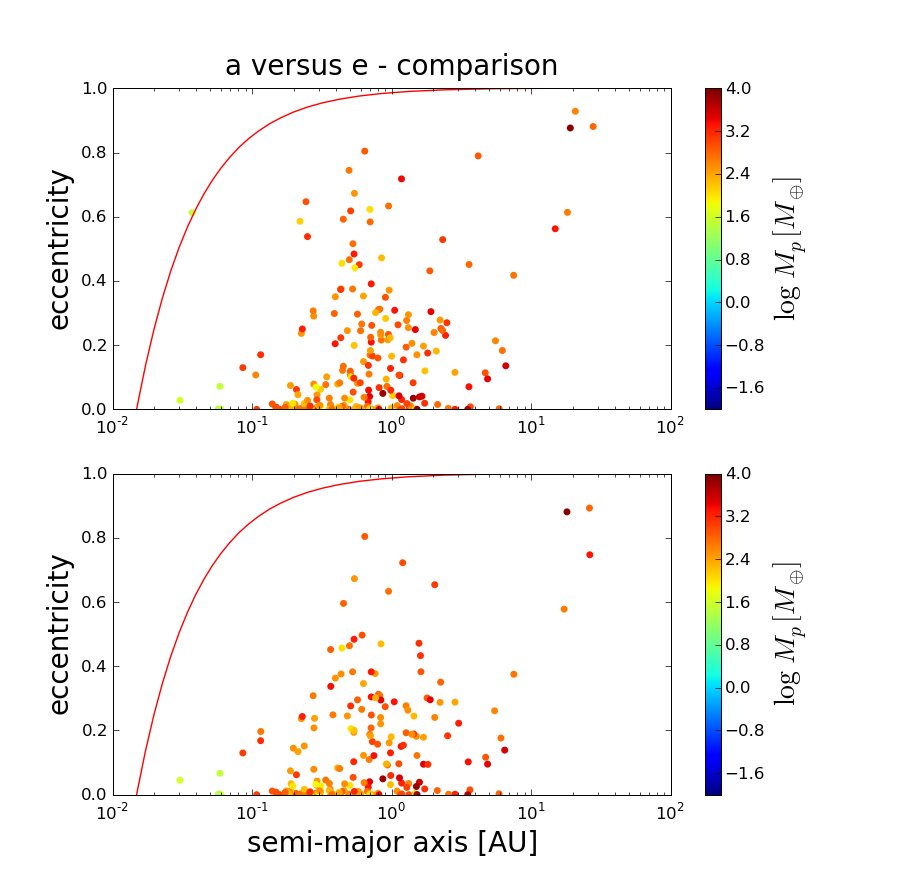}
\caption{Same as Fig. \ref{fig:CD2156_ae_both} for the population with reduced eccentricity damping (RD).}\label{fig:CD2161_ae_both}
\end{figure}

\begin{figure}
\includegraphics[width=0.5\textwidth]{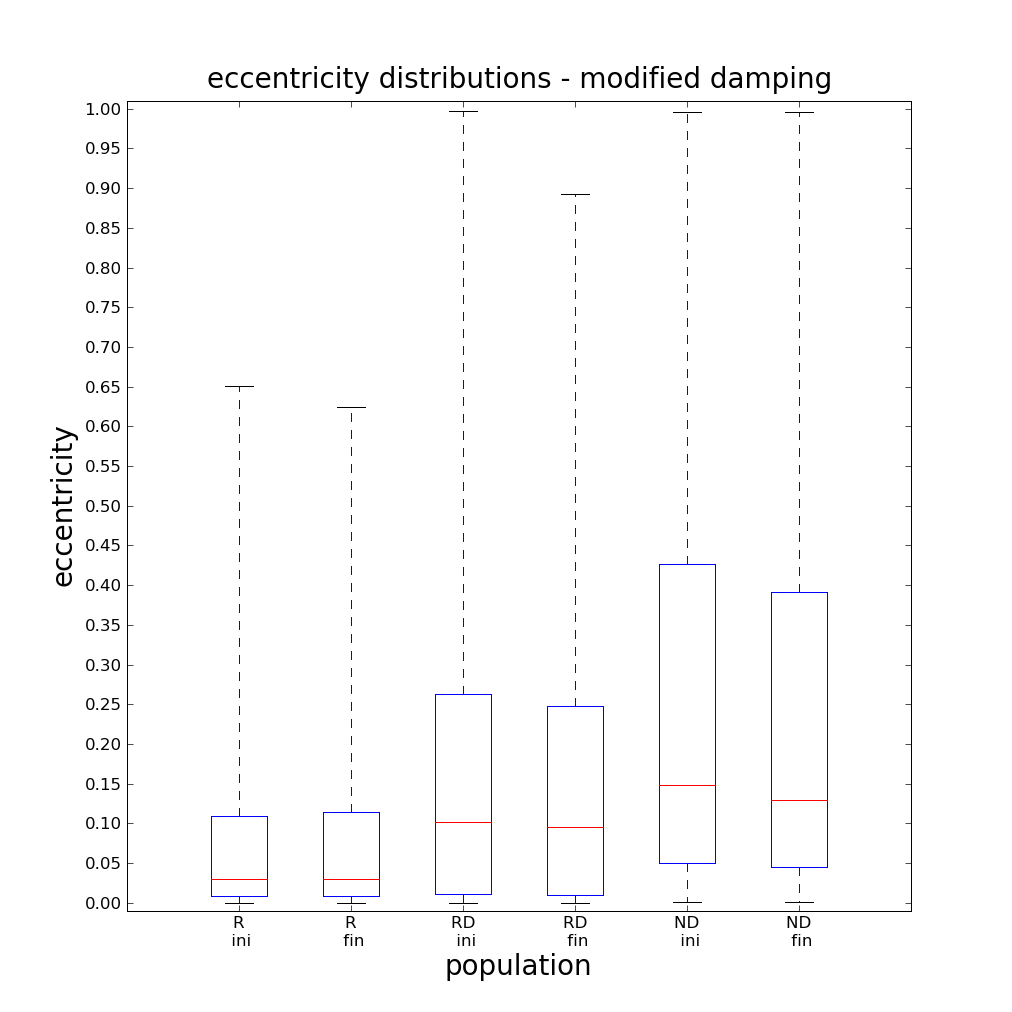}
\caption{Comparison of the eccentricity distributions of the populations R, RD and ND, limited to planets with $K\gsim10\meter\second^{-1}$. The red lines correspond to the median values, the blue boxes respectively denote the interquartile ranges while the whiskers correspond to the minimum and maximum values.}\label{fig:box_CD_damp_10ms}
\end{figure}

\subsection{Stochastic perturbations}

A13 found that the fraction of planet pairs with period ratios close to the nominal locations of low-order mean motion resonances are significantly larger than for observed exoplanets. As resonances can stabilize otherwise unstable systems and thus have an important effect on the eccentricit\sout{i}y evolution of planetary systems, stochastic perturbations to orbital elements from e.g. density fluctuations in the disk can perturb planets out of resonance and thereby increase the likelihood of strong dynamical interaction (scatterings and collisions). To evaluate the effect of small stochastic perturbations, we computed the post-formation evolution for two modified versions of the reference population (R). In the first case, we added a \textit{post facto} perturbation $\delta a\in\lb -0.01,0.01\rb a$ to the semi-major axis of each planet at the end of their formation but before starting the post-formation evolution (population RA). In the second case, we ran the formation code and added a random, uniformly distributed inclination $i\in\lb0\deg,1\deg\rb$ for each planet reaching an inclination of $i\leq 1\deg$ during the formation phase while the gas disk was present (population RI). Contrary to the semi-major axis perturbation, the population with the perturbed inclination is self-consistent, i.e. the perturbation effects were applied to the full formation process rather than just perturbing the final orbital elements. 
The population RA has the same median eccentricities as the reference population for planets with $\rv$, with medians $e=0.030$ at the time of disk dispersal and $e=0.029$ after $100\Myr$. For the full population, the median eccentricity after $100\Myr$ is slightly larger than in the reference population, with $e\simeq2\cdot10^{-3}$ versus $e=1.2\cdot10^{-3}$. Similarly, the eccentricities for the population with perturbed inclinations (RI) are comparable to the eccentricites in the reference population. For planets with $\rv$, the median eccentricity after $100\Myr$ is even slightly lower than in the reference population, with $e=0.028$ versus $e=0.029$. We conclude that the stochastic perturbations we applied to planetary inclinations or semi-major axes have negligible influence on the final eccentricities of planet populations.

\subsection{Comparison with observations}\label{sec:compecc}

\begin{figure}
\includegraphics[width=0.5\textwidth]{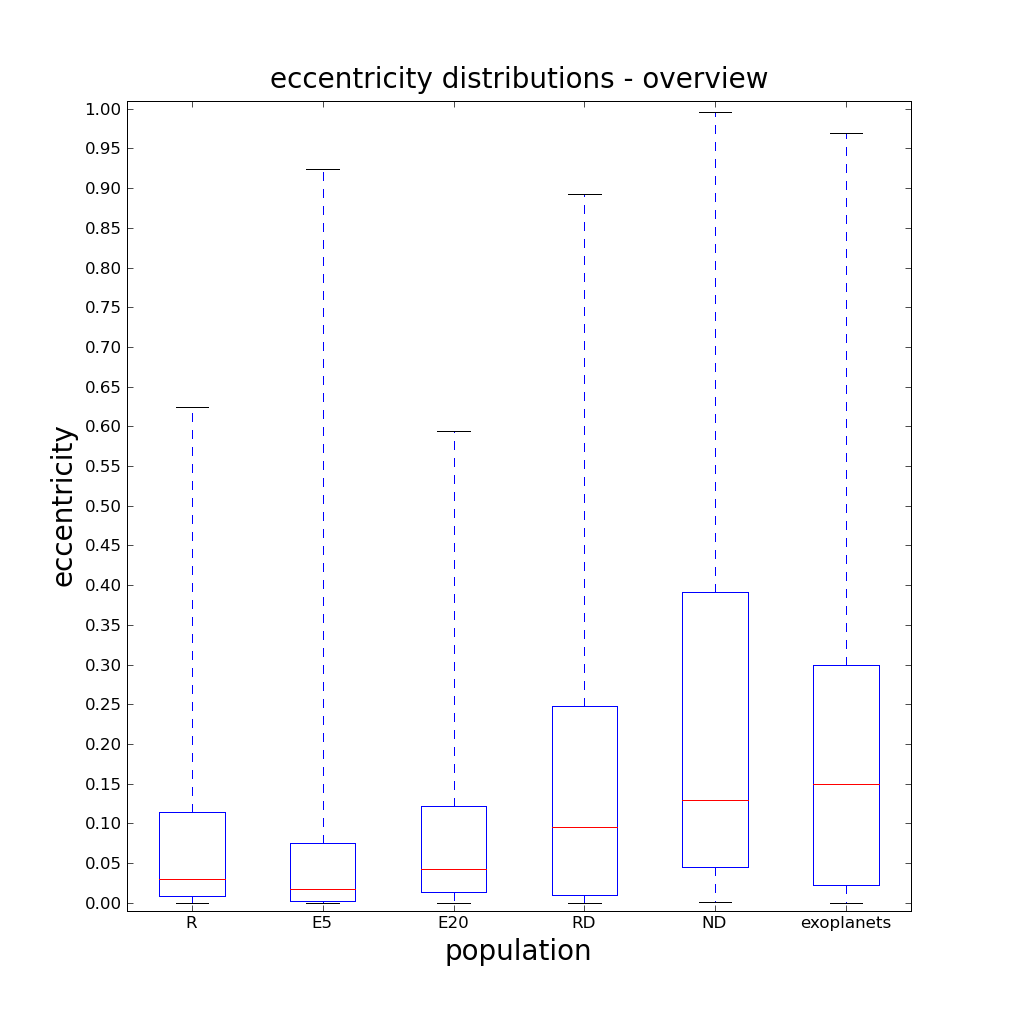}
\caption{Comparison of the eccentricity distributions of the different populations of planets with $K\gsim10\meter\second^{-1}$ after $100\Myr$. The rightmost population corresponds to observed exoplanets. The red lines correspond to the median values, the blue boxes respectively denote the interquartile ranges while the whiskers correspond to the minimum and maximum values.}\label{fig:ecc_boxplot}
\end{figure}

As already stated in A13, no attempt was made to have the synthetic systems reproduce the observed ones. Hence, we did not expect the evolved synthetic population's characteristics to match the observed one either. Nevertheless, we compared the eccentricity distributions of the simulated planet populations to the eccentricity distribution of observed exoplanets\footnote{We take the exoplanet data from \url{www.exoplanets.eu}, a catalog taking a conservative approach to the inclusion of planet data, see \cite{Schneider2011}.} (see Figure \ref{fig:ecc_boxplot}), limiting all samples to planets with $K\geq10\meter\second^{-1}$. We find that the eccentricities in the populations with full eccentricity and inclination damping are significantly smaller than for observed exoplanets. The population where the eccentricity damping timescale was increased by a factor of 10 compared to the nominal value used provides the best agreement\footnote{While the population without damping is in better agreement, the assumption of no damping is very unlikely. We therefore excluded the ND population for this argument.} with observed exoplanet eccentricities. Nevertheless, the median eccentricity of $e=0.095$ after $100\Myr$ is still smaller than the median eccentricity of observed exoplanets ($e=0.15$). A possible source of this difference is the inclusion of planets around stars with no further detected planetary mass companions, but possible distant stellar companions, in the sample of exoplanets. Excluding all exoplanets in systems with multiple stellar components (list taken from \cite{Roell2012}), however, still yields a median eccentricity of $e=0.14$ and does not appreciably change the eccentricity distribution except for the removal of a few highly eccentric planets such as e.g. \textit{HD 20782 b} (see \cite{HD20782b}). A possible source of additional eccentricity not yet included in our model are perturbations by stellar fly-bys, which can increase planetary eccentricities both by direct perturbations and by destabilizing planetary systems, thus increasing the incidence of strong dynamical interactions, see e.g. \cite{Malmberg2011}.

\section{Results: post-formation evolution effects}\label{sec:ltev}

In addition to the change in planetary eccentricities, the post-formation also potentially affects the distributions of other quantities. We compared the mass, semi-major axis and period ratio distributions as well as the chemical composition of planets at the time of disk dispersal and after $100\Myr$.
\\
The fraction of planets removed by the post-formation evolution strongly depends on both the number of embryos and the efficiency of eccentricity and inclination damping. In the reference population, $\sim11\%$ of planets present at the end of the formation phase are removed by collisions and ejections. For the population E5, this amount drops to $<2\%$ while for E20, it increases to $\sim30\%$ mostly due to a significant increase in the number of collisions compared to the reference population. Reducing the efficiency of eccentricity and inclination damping by a factor of 10 (population RD) results in a slight increase in the fraction of removed planets ($\sim13\%$) through an increase in the fraction of ejected planets ($\sim3\%$ for RD versus $<1\%$ for R), while completely removing the eccentricity and inclination damping results in $\sim23\%$ of all planets removed through collisions and ejections, with marked increase in both types of removal. The situation is similar for the fraction of systems experiencing collisions and/or ejections in each population (see Table \ref{tab:popov_eve}), with a sharp increase for increasing number of embryos and an increase for decreasing damping. Any change in the distribution of mass, semi-major axis or period ratios is therefore likely most pronounced in the populations E20 and ND. The addition of stochastic perturbations results in similar fractions of lost planets and active systems in the case of inclination perturbations (RI), but in substantially different fractions for perturbations of the semi-major axes (RA). For the latter, the fraction of lost planets increases by almost a third to $f_{\text{lost}}=16\%$, while the fraction of active systems, i.e. systems with either collisions or ejections, increases from $f_{\text{sys,active}}=36\%$ in the reference population to $f_{\text{sys,active}}=57\%$ in the perturbed population. The large increase in activity despite the rather small perturbation suggests that many systems in our reference population are close to instability. Most likely, such instabilities are stabilized by the presence of mean-motion resonances, which are then broken by the semi-major axis perturbations. The lack of a significant increase in eccentricity for the perturbed population, however, implies that the increased activity predominately affects the lower-mass planets. 
\begin{table*}[htbp]\caption{Overview of planet removal. $n_p$ denotes the number of planets present after the formation phase, $n_{sys}$ denotes the number of systems in the population, $f_{lost}$ denotes the fraction of planets removed by ejections and collisions during the post-formation evolution, and $f_{sys,active}$ denotes the fraction of systems experiencing collision, ejections or both. The population designations refer to the reference population with 10 embryos and full damping (R), the population with 5 and 20 embryos (E5, E20), the population with no (ND) and reduced eccentricitiy and inclination damping (RD), and the population with a stochastic perturbation to the semi-major axes (RA) or to the inclinations (RI).}\label{tab:popov_eve}
\centering
\begin{tabular}{|c|c|c|c|c|c|c|c|} \hline
 & R & E20 & E5 & ND & RD & RA & RI \\ \hline
$n_{sys}$ & 344 & 228 & 483 & 264 & 343 & 344 & 413 \\ \hline
$n_p$ & 2834 & 3241 & 2171 & 1853 & 2480 & 2834 & 3309 \\ \hline
$f_{lost}$ & 0.11 & 0.30 & 0.02 & 0.23 & 0.13 & 0.16  & 0.12 \\ \hline
$f_{sys,active}$ & 0.36 & 0.78 & 0.06 & 0.67 & 0.41 & 0.57 &  0.38 \\ \hline
\end{tabular}

\end{table*}

\subsection{The planetary mass function}

\begin{figure}
\includegraphics[width=0.5\textwidth]{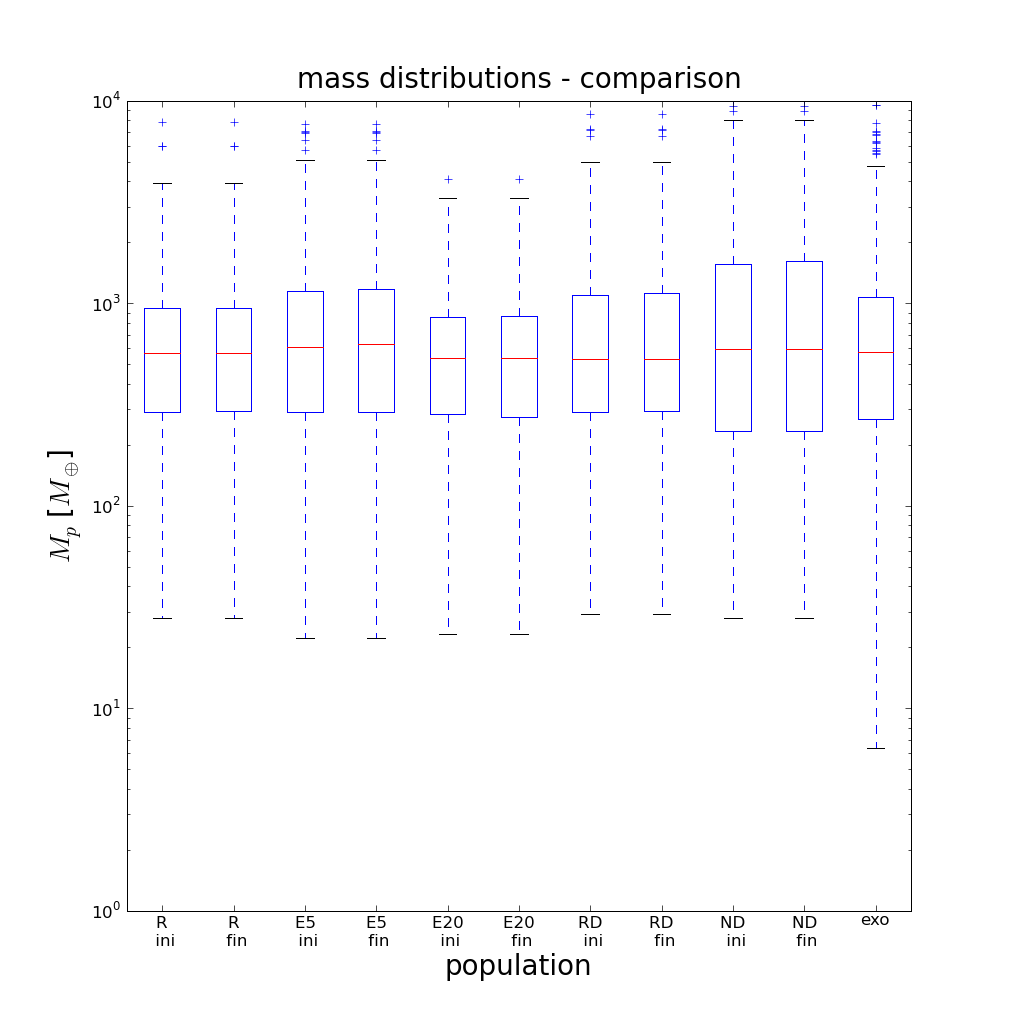}
\caption{Comparison of the mass distributions for different populations of planets with $K\gsim10\meter\second^{-1}$. The red lines correspond to the median values, the blue boxes respectively denote the interquartile ranges while the whiskers correspond to $5$ times the interquartile range, values outside this range are denoted by a blue $+$-sign.}\label{fig:pmf_boxplot}
\end{figure}

\begin{figure}
\includegraphics[width=0.5\textwidth]{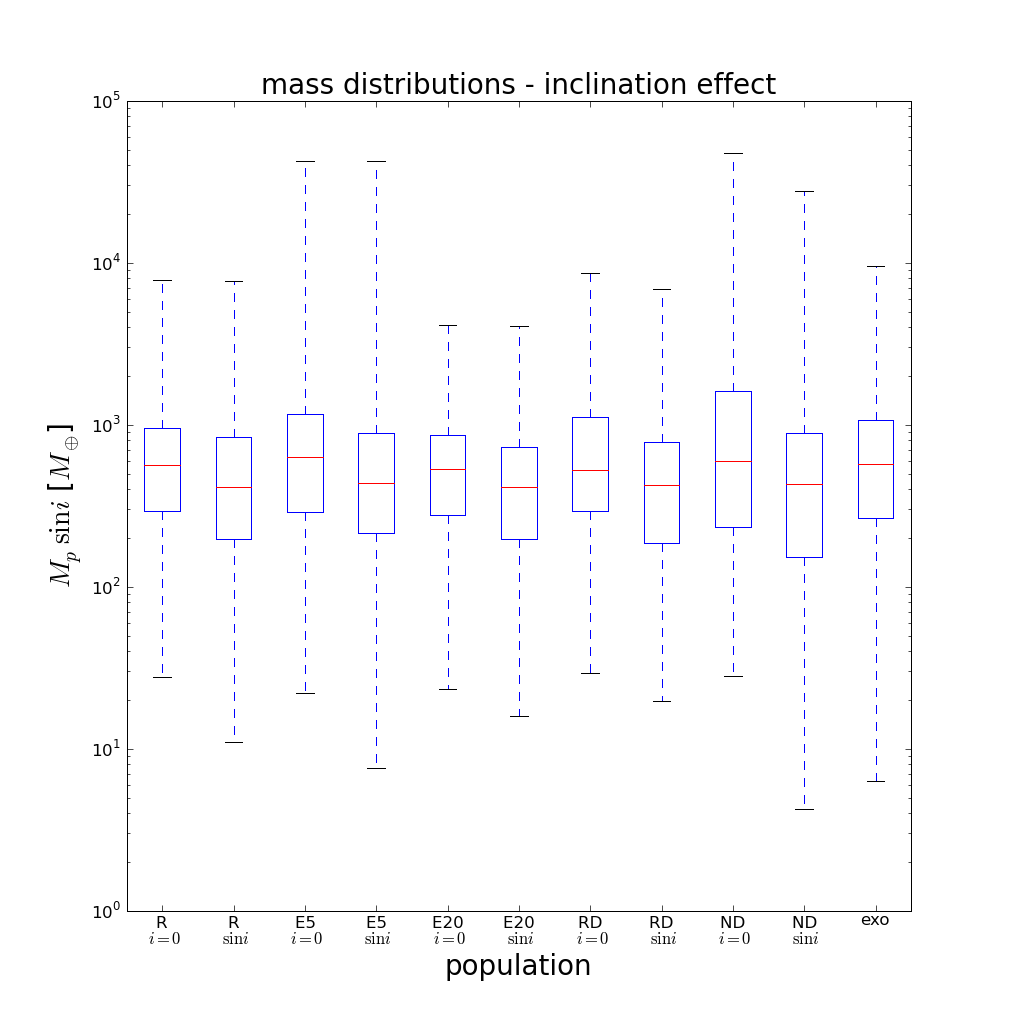}
\caption{Comparison of the $M\sin i$ distributions for different populations of planets. The red lines correspond to the median values, the blue boxes respectively denote the interquartile ranges while the whiskers correspond to the maximum and minimum masses.}\label{fig:msinibox}
\end{figure}

The effect of the post-formation on mass distributions of the populations studied is negligible for planets with $K\geq10\meter\second^{-1}$ (see Figure \ref{fig:pmf_boxplot}), with both the distributions at the time of disk dispersal and after $100\Myr$ in solid agreement with the mass distribution of observed exoplanets when taking the observed exoplanet masses to be their true masses. For exoplanets detected by radial velocity measurements, however, planetary masses are in general only known up to a factor $\sin i$, $i$ being the angle between the observer and the orbital plane of the planet. Including a random line-of-sight inclination drawn from $i\sim\sin i$ (see e.g. \cite{Ho2011}) to compute the $m\sin i$, we find that our planets have a smaller median $m\sin i$ than observed exoplanets (see Fig. \ref{fig:msinibox}). Moreover, using a two-sample Kolmogorov-Smirnov test, all populations studied show a better agreement with the observed mass distribution for the true population rather than for the $\sin i$ corrected distribution (cf. Tab. \ref{tab:ks_mass}). Indeed, the null hypothesis that the $m\sin i$ are drawn from the same distribution as observed exoplanet $m\sin i$ cannot be rejected only for the population E5. Comparing the true masses to observed $m\sin i$, the null hypothesis can only be rejected for the population ND, whereas the remaining populatios are in solid agreement with the observed distribution. As $m\geq m\sin i$, the (massive) planets in our populations appear to be somewhat less massive overall than observed exoplanets. 
In addition to the planets remaining after $100\Myr$, we also tracked the masses of planets ejected both during the formation phase and the subsequent post-formation evolution. Microlensing surveys claim a large number of \textit{free-floating} planets with $m\sim\MJ$, with e.g. \cite{Sumi2011} finding that \textit{free-floating} planets in the jovian mass range are about twice as common as main-sequence stars. Even though the planetary mass functions in our simulations are comparable to the planetary mass function for observed exoplanets, our simulations generally produce only few ejections (see above) and an even lower fraction of ejected planets with masses $m\gsim100\Me$. Of all planets with $m\geq100\Me$, only $\lsim7\%$ are ejected from simulations with the full eccentricity and inclination damping (populations R, E5, E20 and RA), and $\sim19\%$ and $\sim43$, respectively, for the population with reduced (RD) and without (ND) eccentricity damping. With the fraction of planets with $m\geq100\Me$ between $5\%$ and $10\%$ in our simulations, ejections during the formation phase and the post-formation evolution alone cannot explain a large population of jovian-mass \textit{free-floating} planets.
\begin{table*}[htbp]\caption{$p$-values of two-sided Kolmogorov-Smirnov tests for the planets with $\rv$ in each population. The first column compares the true mass distribution with the distribution including the $\sin i$ correction, the second column compares the true mass distribution with the observed mass distribution of exoplanets with $\rv$, and the last column compares the mass distribution including the $\sin i$ correction with the observed mass distribution.}\label{tab:ks_mass}
\centering
\begin{tabular}{|c|c|c|c|} \hline
population &  $m$ vs $m\sin i$ & $m$ vs observed & $m\sin i$ vs observed \\ \hline
R &  0.051 & 0.523 & 0.009 \\ \hline
E5 & 0.041 & 0.553 & 0.193 \\ \hline
E20 & 0.041 & 0.246 & 0.002 \\ \hline
RD & 0.003 & 0.556 & 4.84$\cdot10^{-5}$ \\ \hline
ND & 0.269 & 0.021 & 0.006 \\ \hline
RA & 0.084 & 0.548 & 0.043 \\ \hline
RI & 0.046 & 0.741 & 0.001 \\ \hline
\end{tabular}
\end{table*}
\begin{figure}
\includegraphics[width=0.5\textwidth]{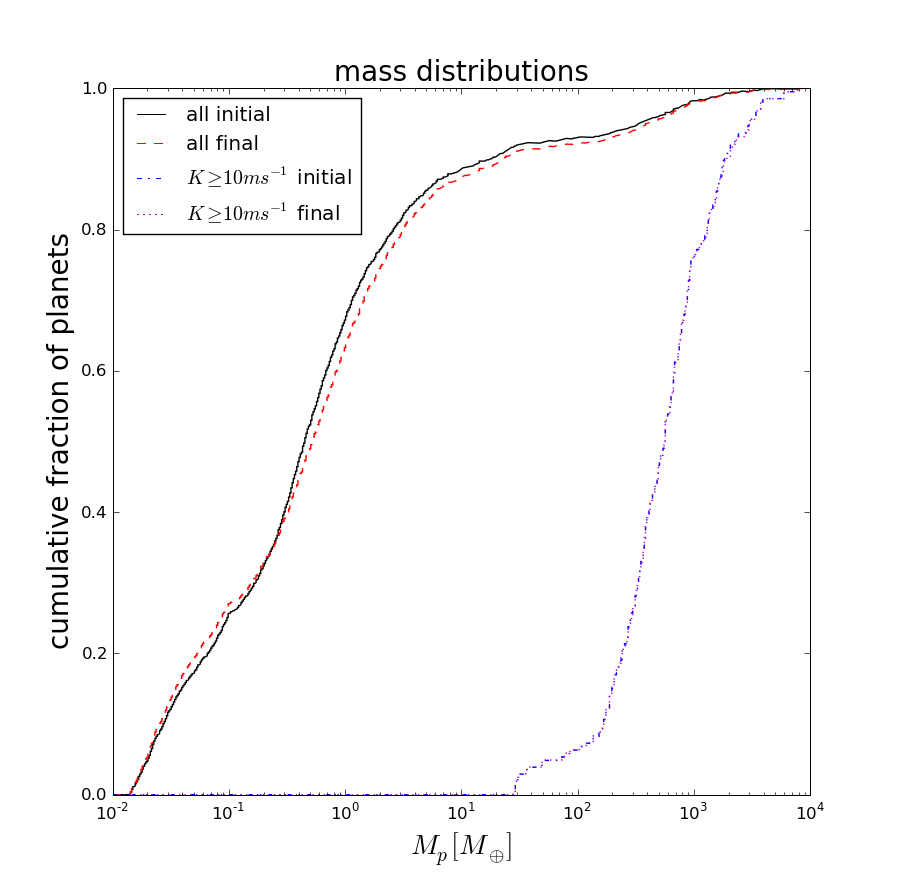}
\caption{Planetary mass functions for the reference population (R). The black solid line and the red dashed line respectively correspond to the distributions at the time of disk dispersal and after $100\Myr$, the blue dash-dotted line and the magenta dotted line respectively correspond to the distributions at the time of disk dispersal and after $100\Myr$ restricted to planets with $K\geq10\meter\second^{-1}$.}\label{fig:CD2133_pmf_both}
\end{figure}
\begin{figure}
\includegraphics[width=0.5\textwidth]{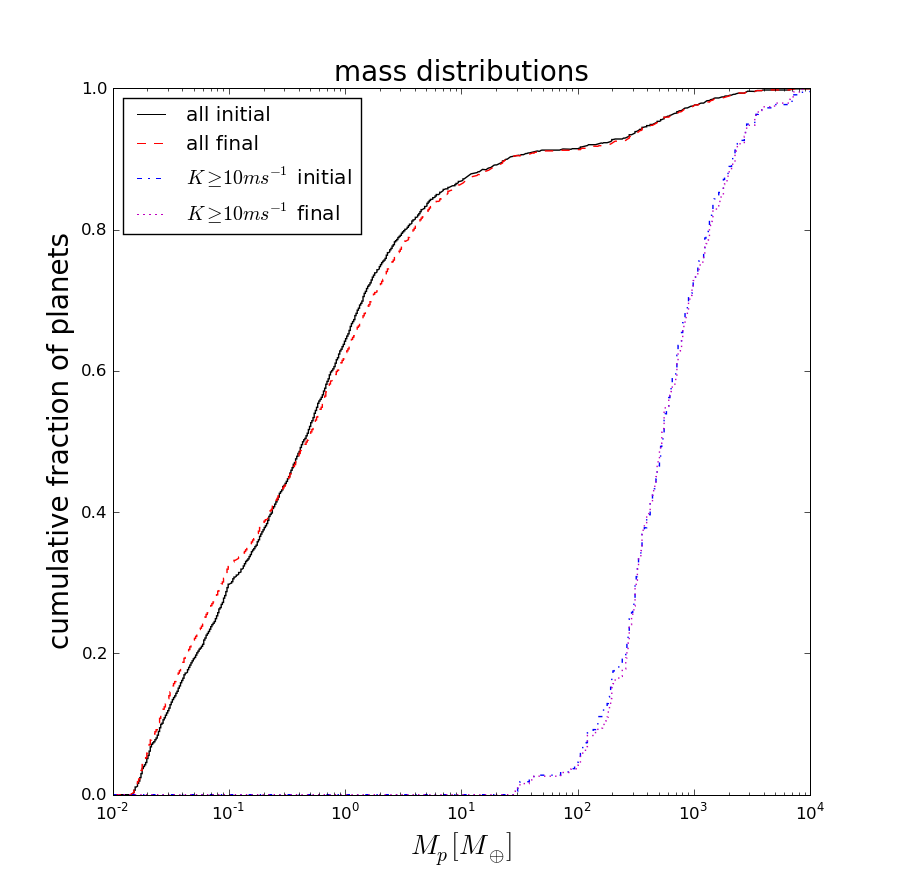}
\caption{Same as Fig. \ref{fig:CD2133_pmf_both} for the population with reduced eccentricity damping (RD).}\label{fig:CD2161_pmf_both}
\end{figure}

\begin{figure}
\includegraphics[width=0.5\textwidth]{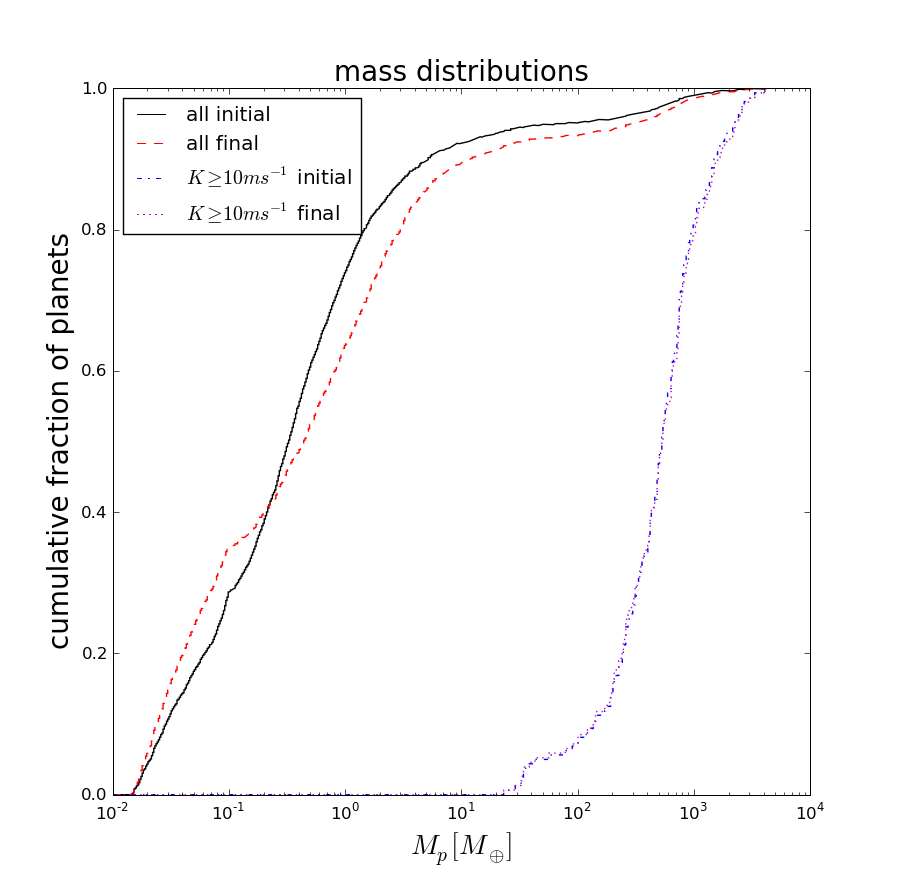}
\caption{Same as Fig. \ref{fig:CD2133_pmf_both} for the population with 20 initial embryos (E20).}\label{fig:CD2160_pmf_both}
\end{figure}

\begin{figure}
\includegraphics[width=0.5\textwidth]{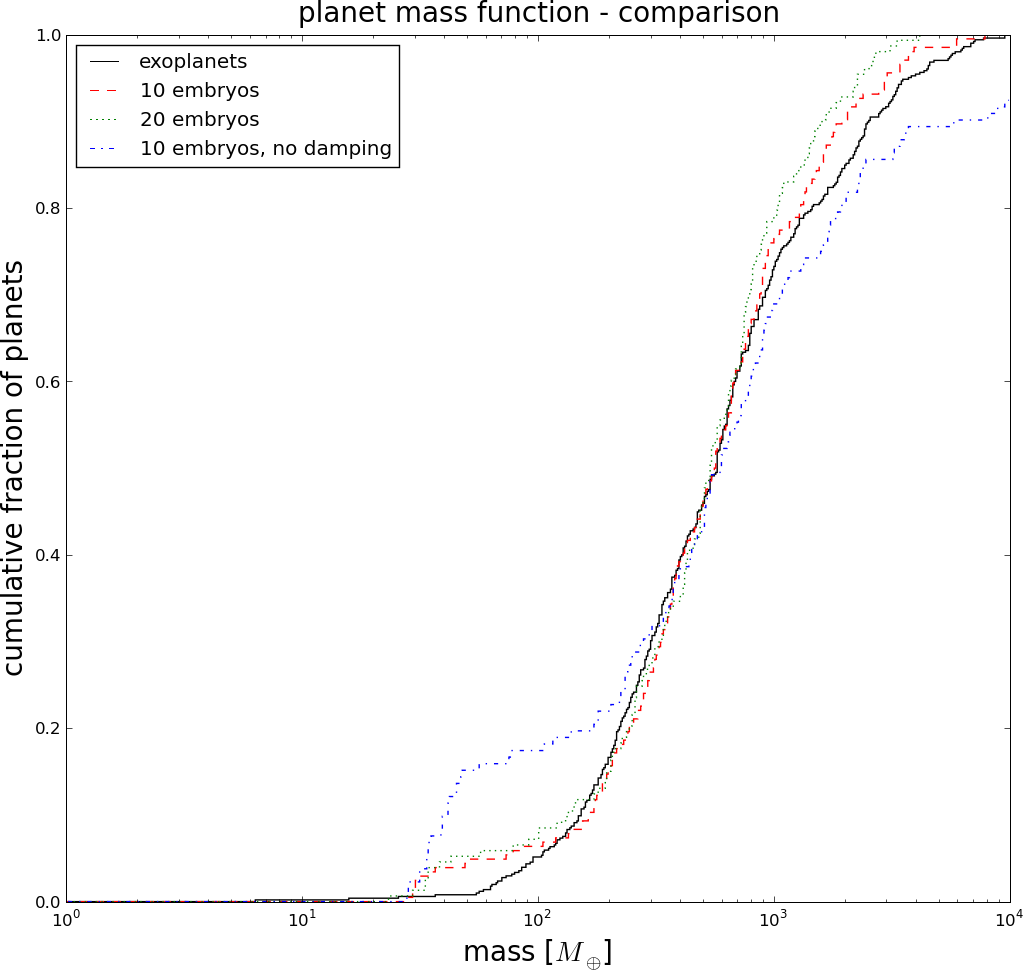}
\caption{Comparison of the cumulative mass distributions after $100\Myr$ for different populations of planets with $K\gsim10\meter\second^{-1}$. The black solid line denotes the distribution for observed exoplanets, the red dashed line for the reference population (R), the blue dot-dashed line for the population without eccentricity and inclination damping (ND) and the green dotted line for the population with 20 intial embryo seeds (E20).}
\end{figure}

\subsection{Semi-major axis distribution}

\begin{figure}
\includegraphics[width=0.5\textwidth]{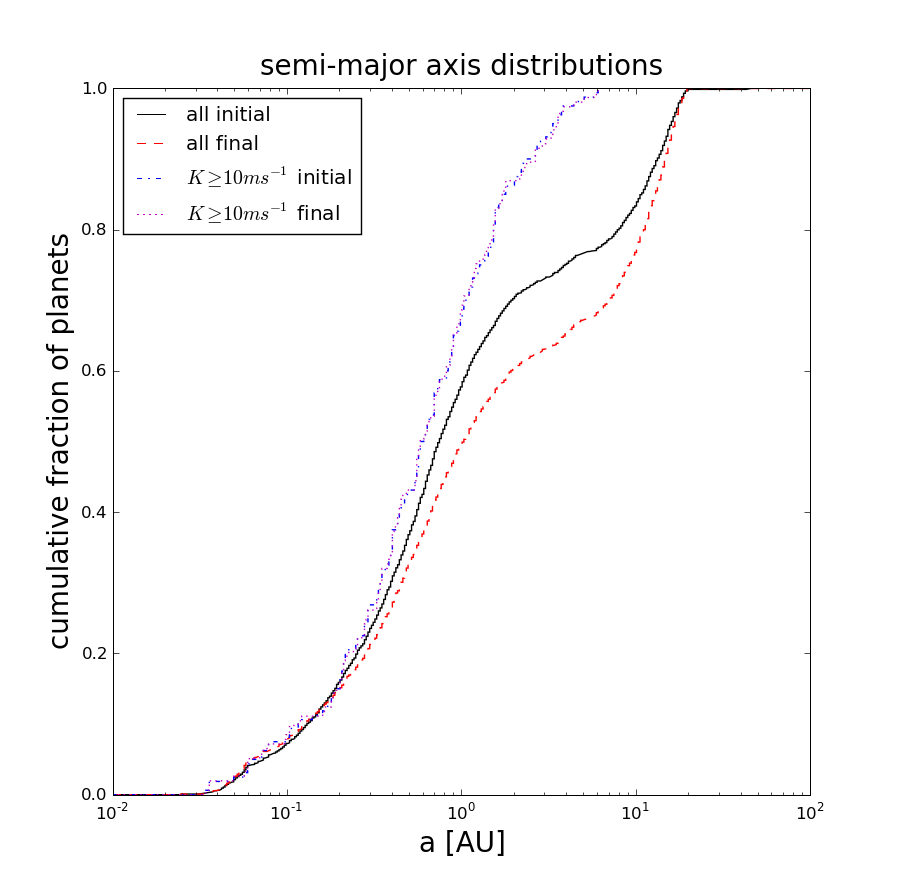}
\caption{Semi-major axis distribution for all planets in the population with a larger number of initial embryo seeds (E20). The black solid line and the red dashed line respectively correspond to the distributions at the time of disk dispersal and after $100\Myr$, the blue dash-dotted line and the magenta dotted line respectively correspond to the distributions at the time of disk dispersal and after $100\Myr$ restricted to planets with $K\geq10\meter\second^{-1}$.}\label{fig:ahist}
\end{figure}

Similarly to the mass distributions, the distributions of the semi-major axes remain almost unchanged by the post-formation evolution for planets with $K\gsim10\meter\second^{-1}$. For the full populations, the changes increase with increasing number of embryos and with decreasing eccentricity damping, with the population with 20 initial embryo seeds exhibiting the largest changes to the semi-major axis distribution (Figure \ref{fig:ahist}). With the number of scattered planets being smaller than the number of planets undergoing collisions, the changes in the distributions are an effect of removing a part of the population rather than of changing the semi-major axes of the planets. The latter is also evidenced by the lack of change in the distribution of period ratios (see Section \ref{sec:pr}).

\subsection{The chemical composition of planets}

\begin{figure}
\includegraphics[width=0.5\textwidth]{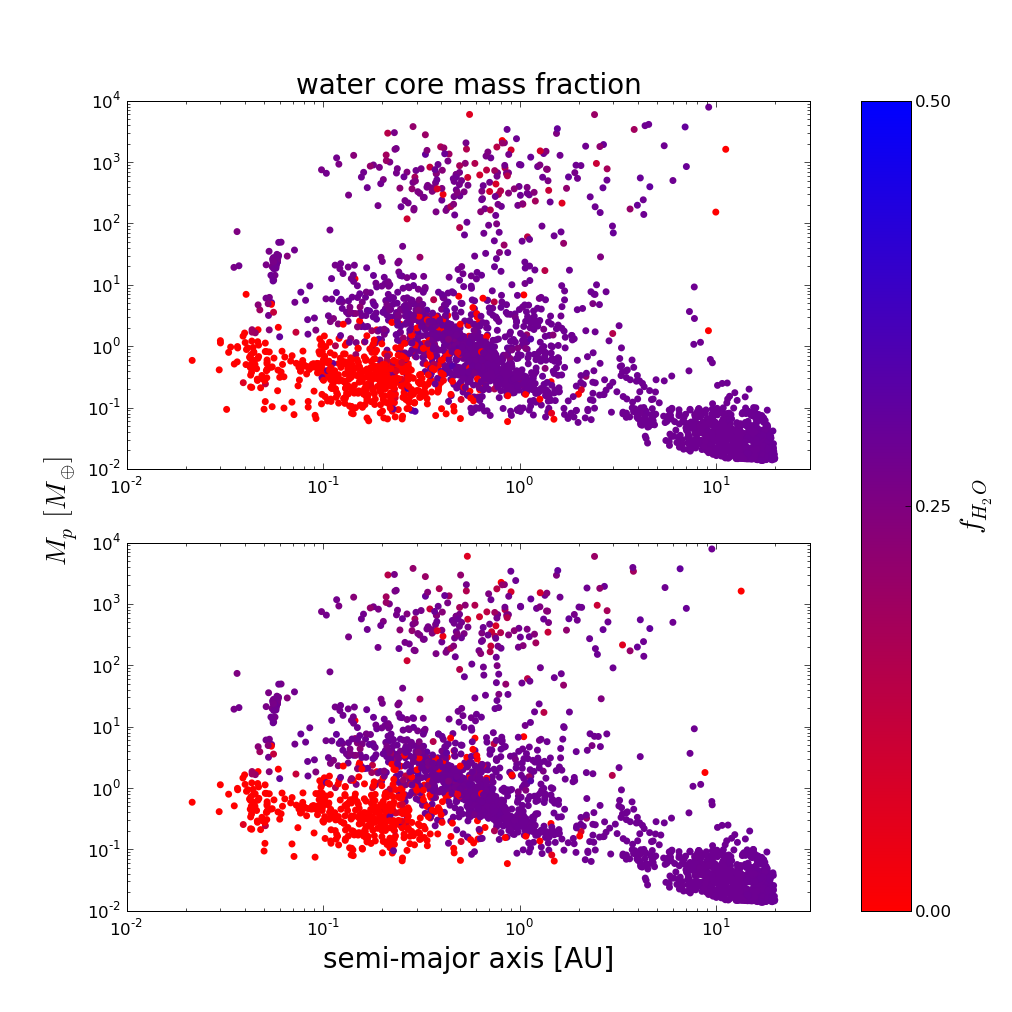}
\caption{Initial (upper panel) and final (lower panel) water mass fractions $\fwat$ for all planets in the reference population (R) as a function of semi-major axis and mass.}\label{fig:wfR}
\end{figure}

\begin{figure}
\includegraphics[width=0.5\textwidth]{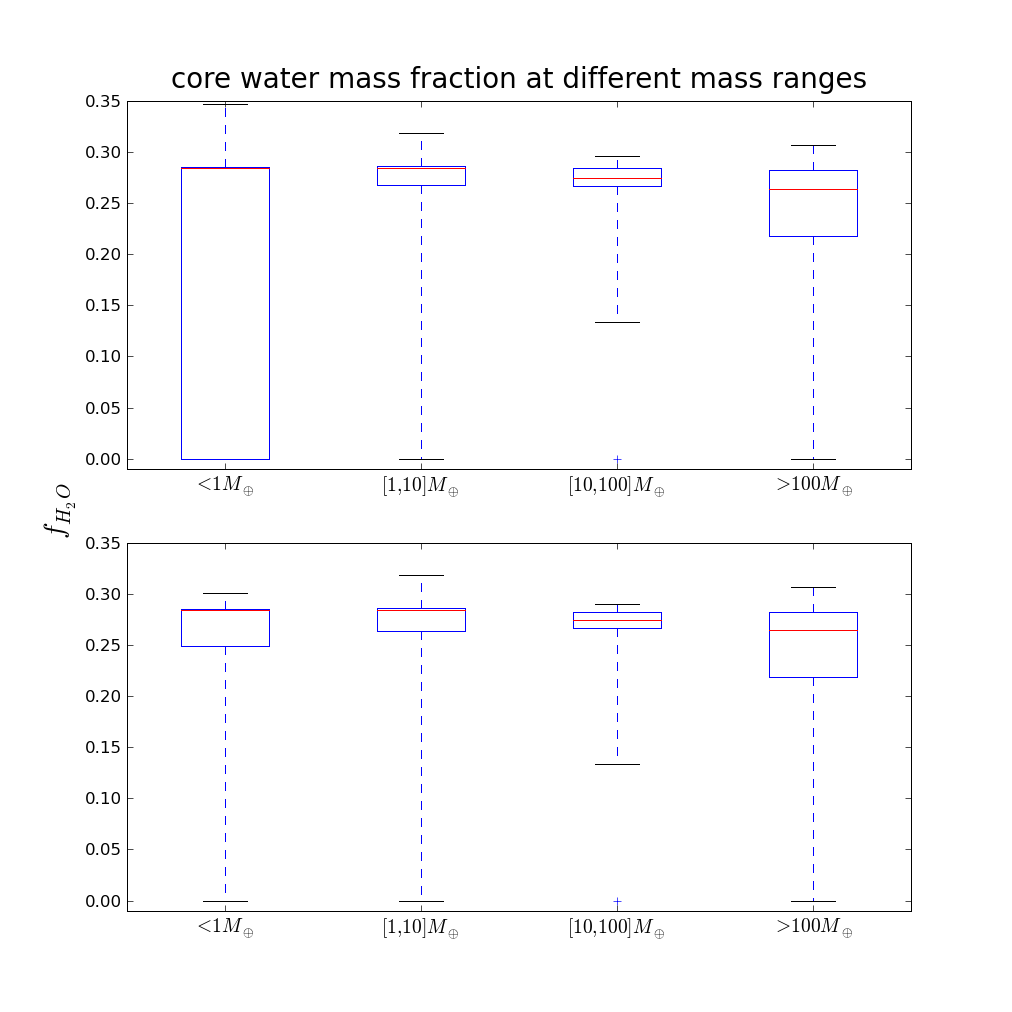}
\caption{Initial (upper panel) and final (lower panel) water mass fraction $\fwat$ distributions for different mass bins.}\label{fig:wfRm}
\end{figure}

\begin{figure}
\includegraphics[width=0.5\textwidth]{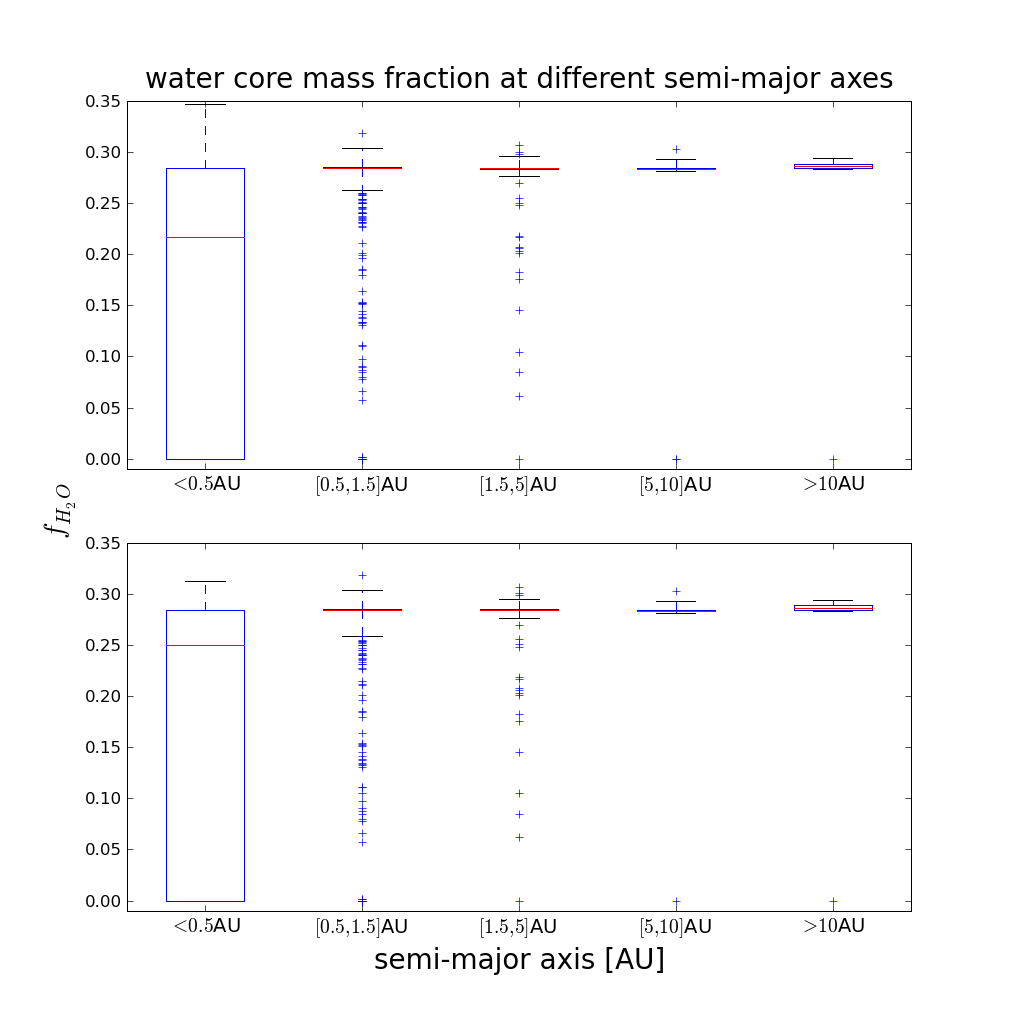}
\caption{Initial (upper panel) and final (lower panel) water mass fraction $\fwat$ distributions for different semi-major axis bins.}\label{fig:wfRa}
\end{figure}

For the same disk parameters and evolutions as in the planet formation code, \cite{Marboeuf2014a,Marboeuf2014b} and \cite{Thiabaud2014} respectivly computed the distribution of volatile and refractory elements from chemical condesation sequences, thus allowing for the computation of the chemical compositions of planets in the Bernese model. Using the initial compositions computed by \cite{Marboeuf2014a,Marboeuf2014b} and \cite{Thiabaud2014} in the case of a non-irradiated disk\footnote{The inclusion of disk irradiation affects both the condensation sequence and the formation and evolution of planets. Simulations including these effects self-consistently will be the topic of future work, while their impact on disk chemistry is studied in \cite{Thiabaud2014}.}, we computed the chemical compositions after $100\Myr$ for the reference population and the population with an increased number of embryos by following the collisional evolution of the planets.\\
Comparing the initial and final distribution of mass fractions of water for the planetary cores $\fwat$ (see Fig. \ref{fig:wfR}), we find that post-formation evolution globally has a minimal impact on $\fwat$ as the population median remains at $\fwat=0.284$. The spread in $\fwat$ decreases somewhat as the lower quartile increases from $\fwat=0.241$ to $\fwat=0.255$, indicating that in our simulation, post-formation evolution reduced the amount of ``dry'' planets. Comparing the evolution of the distribution of $\fwat$ at different mass and semi-major axis ranges (Figs. \ref{fig:wfRm} and \ref{fig:wfRa}, respectively), we find that planets with $m_p\leq1\Me$ and planets with $a_p\leq0.5\AU$ exhibit significant change in $\fwat$ whereas more massive planets and planets on wider orbits show almost no change in $\fwat$. However, for close-in, low-mass planets, the fraction of ``dry'' planets shows a slight increase after $100\Myr$.\\
The lack of change in $\fwat$ for massive planets is primarily due to the very low frequency of collisions (and, to a lesser degree, ejections) involving massive planets. \\
Note that we take into account any potential loss of water due to collisions neither during the formation phase nor during the post-formation evolution.

\subsection{Period ratios}\label{sec:pr}

The distributions of period ratios of planets with $K\geq10\meter\second^{-1}$ remain virtually unchanged by the post-formation evolution for all planet populations we simulated. For the full populations, the post-formation evolution slightly increases the fraction of pairs with large period ratios ($T_{out}/T_{in}\gsim10$). The latter effect appears to be an increasing function of the number of initial embryo seeds: for 5 embryos, the effect of the post-formation is negligible, for 10 embryos, the effect is small, and for 20 embryos, the effect becomes somewhat important. Conversely, the fraction of planet pairs close to low-order mean-motion resonances decreases with increasing number of embryos, both for the full populations and for the populations restricted to planets with $K\geq10\meter\second^{-1}$. Reducing the efficiency of eccentricity and inclination damping by increasing the damping timescale by a factor of 10 does not significantly change the period ratio distributions compared to the reference case (see Figure \ref{fig:CD2161_pr_all}). Completely removing the damping, however, produces a starkly different distribution with no discernable population of pairs in low-order mean motion resonances (Figure \ref{fig:CD2156_pr_all}).\\
The population with a random perturbation to the semi-major axes (population RA), while showing an overally similar distribution of period ratios for planets with $K\geq10\meter\second^{-1}$, has a lower fraction of planets within $\pm1\%$ of the nominal locations of low-order mean motion resonances after $100\Myr$ compared to the reference population. The ratio of the fraction of systems within one percent of the $2:1$ mean motion resonance in the reference population at the end and at the beginning of the post-formation evolution, for instance, is $f_{2:1,fin}/f_{2:1,ini}=0.17/0.18\approx0.94$, whereas the same ratio for the population RA is $f_{2:1,fin}/f_{2:1,ini}=0.10/0.14\approx0.71$ comparable to the values for the population with more initial embryo seeds (E20) which has $f_{2:1}\simeq0.15$ initially and $f_{2:1}\simeq0.11$ after the post-formation evolution. Compared to $f_{2:1}\simeq0.05$ for observed exoplanet pairs, this is still considerably larger and points to our model either lacking strong dynamical interactions which could force the probably resonant planet pairs away from resonance, or having an overly efficient migration producing (too) many pairs in resonance. Both possibilities would also affect the distribution of eccentricities discussed in Sec. \ref{sec:compecc}. 

\begin{figure}
\includegraphics[width=0.5\textwidth]{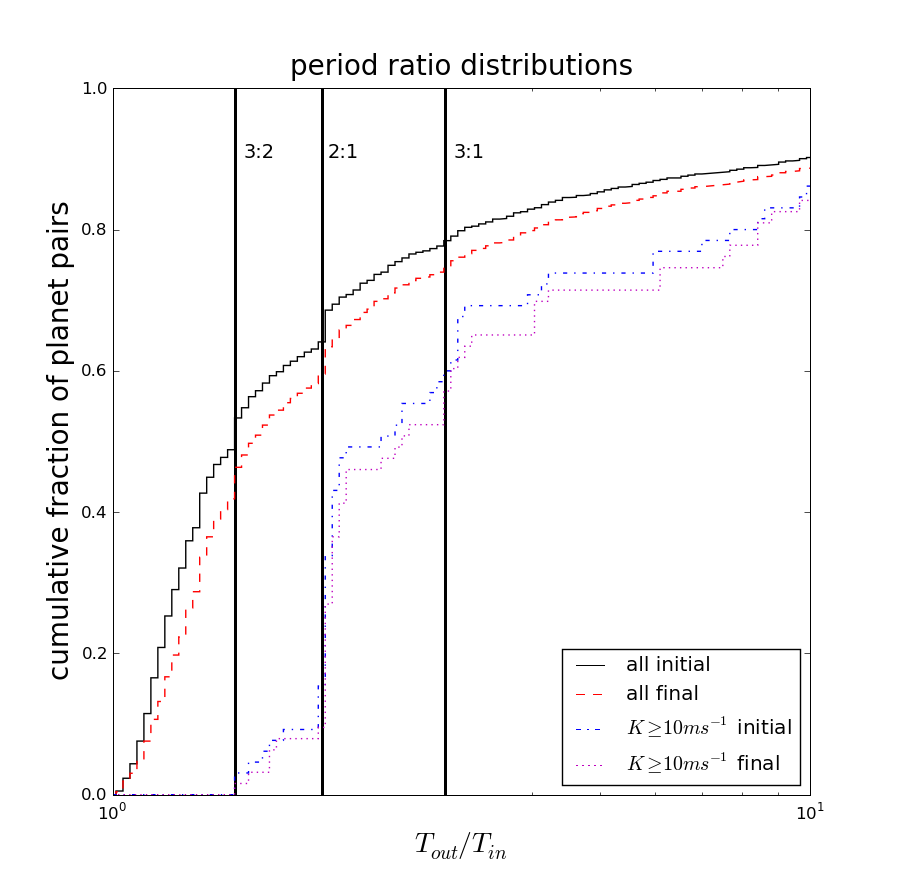}
\caption{Period ratios for planet pairs in the reference population (R) for which both planets have $K\geq10\meter\second^{-1}$. The black, solid line corresponds to the initial distribution for all planet pairs, the dashed, red line to the distribution after $100\Myr$ for all planet pairs. The blue dot-dashed line and the magenta dotted lines respectively denote the initial and final distributions for planet pairs where both planets have $K\geq10\meter\second^{-1}$.}\label{fig:CD2133_pr_10ms_both}
\end{figure}

\begin{figure}
\includegraphics[width=0.5\textwidth]{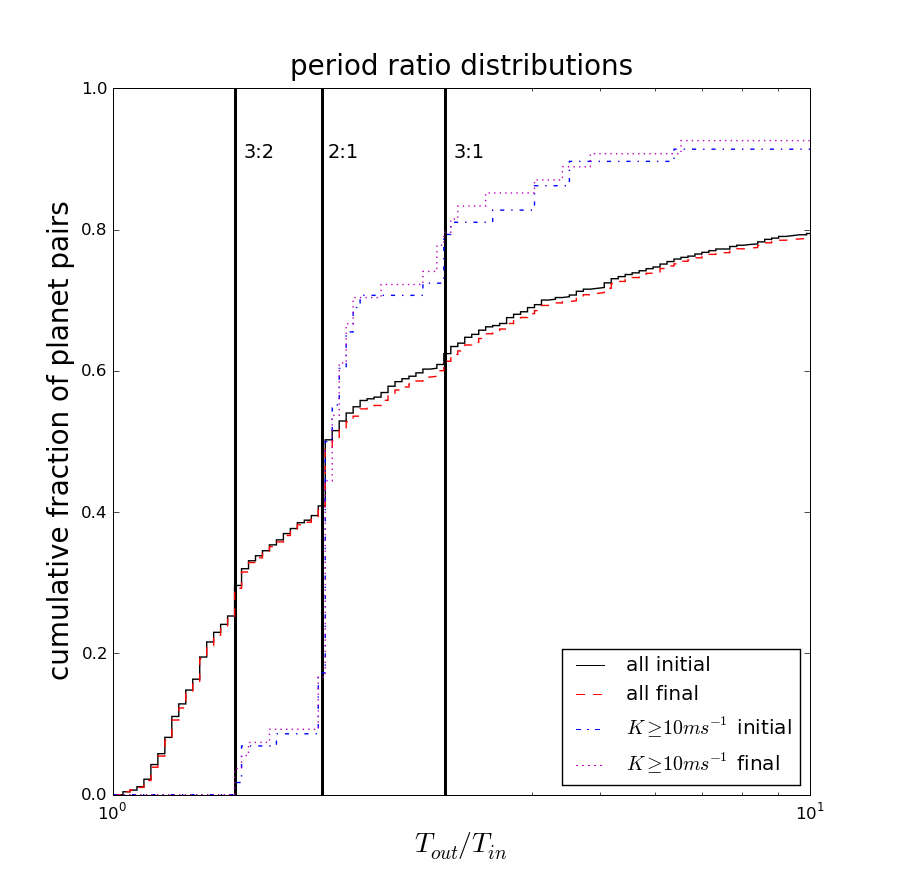}
\caption{Same as Fig. \ref{fig:CD2133_pr_10ms_both} for the population with 5 initial embryos (E5).}\label{fig:CD2159_pr_10ms_both}
\end{figure}

\begin{figure}
\includegraphics[width=0.5\textwidth]{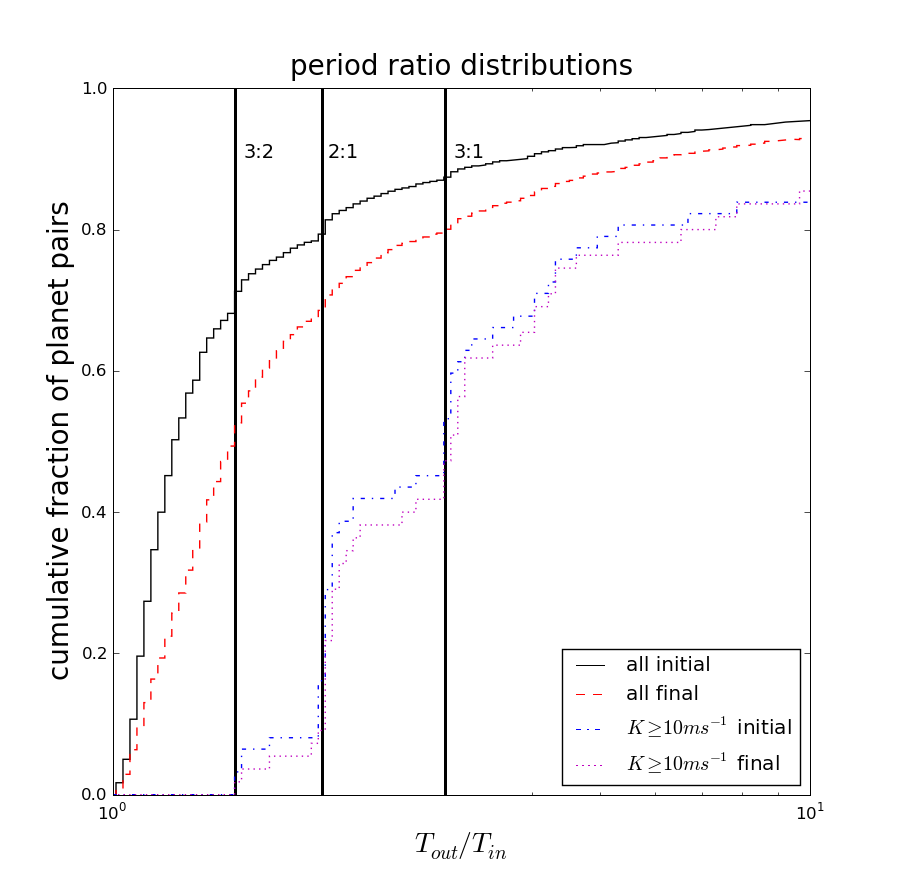}
\caption{Same as Fig. \ref{fig:CD2133_pr_10ms_both} for the population with 20 initial embryos (E20).}\label{fig:CD2160_pr_10ms_both}
\end{figure}

\begin{figure}
\includegraphics[width=0.5\textwidth]{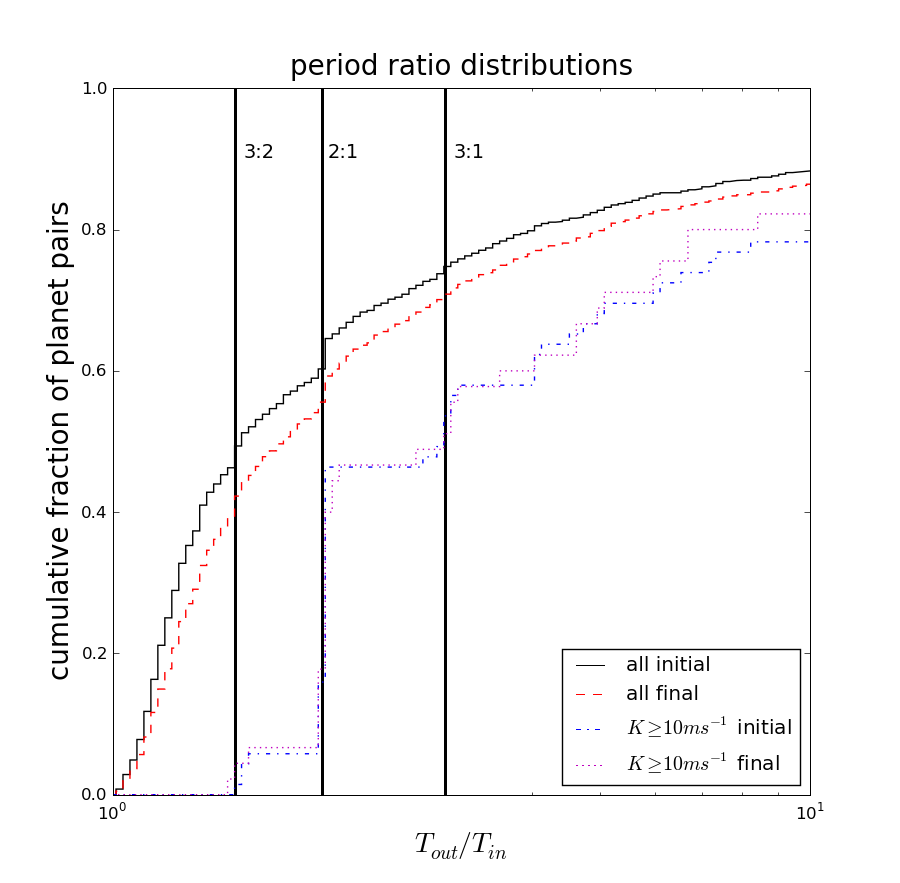}
\caption{Same as Fig. \ref{fig:CD2133_pr_10ms_both} for the population with reduced eccentricity damping (RD).}\label{fig:CD2161_pr_all}
\end{figure}

\begin{figure}
\includegraphics[width=0.5\textwidth]{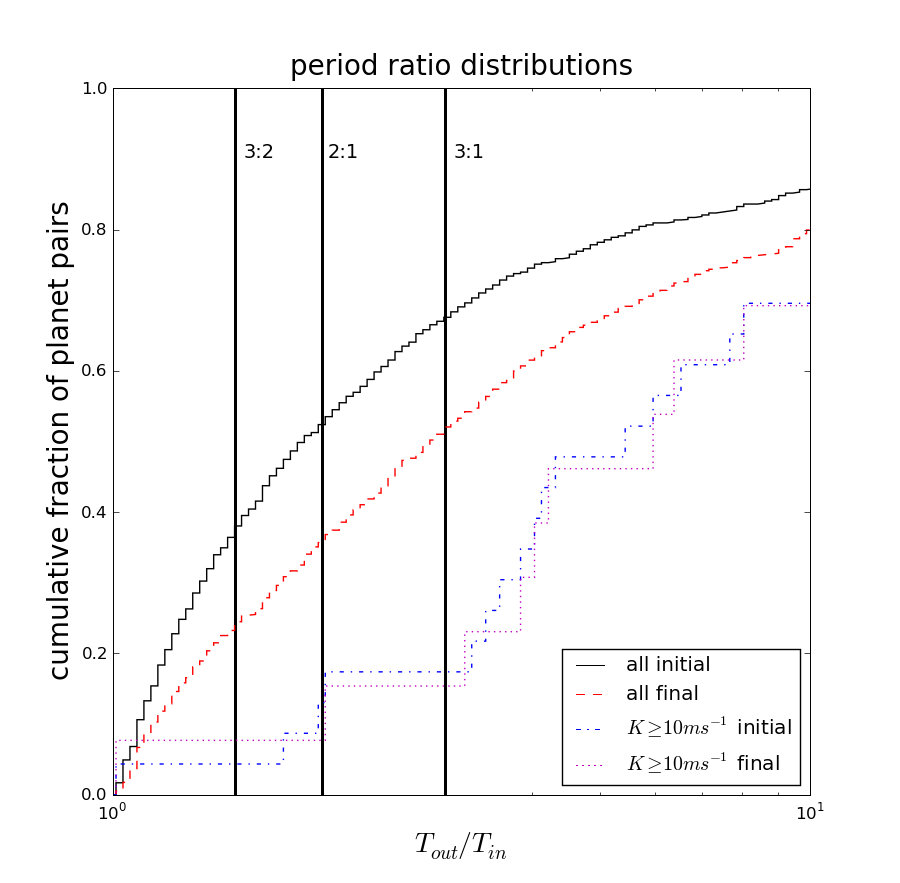}
\caption{Same as Fig. \ref{fig:CD2133_pr_10ms_both} for the population without eccentricity damping (ND).}\label{fig:CD2156_pr_all}
\end{figure}

\begin{figure}
\includegraphics[width=0.5\textwidth]{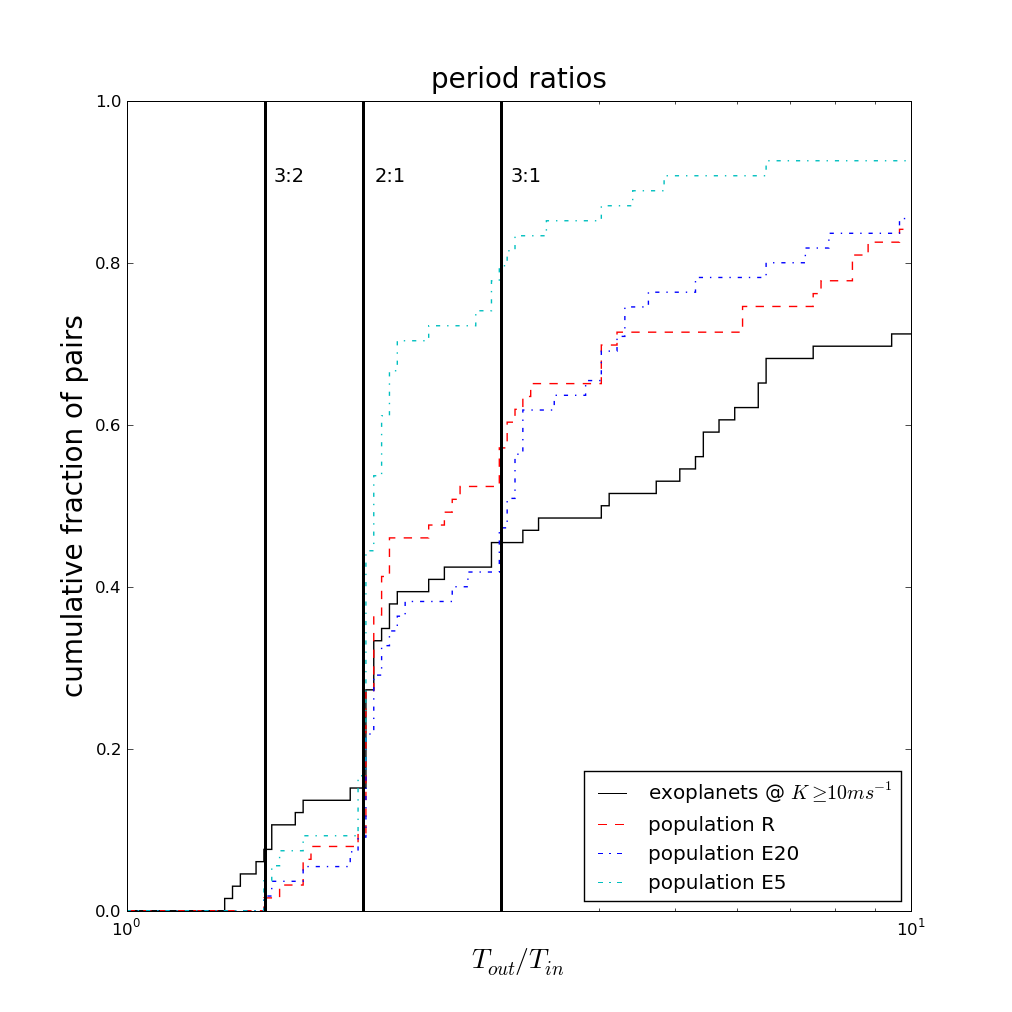}
\caption{Comparison of the period ratios after $100\Myr$ for different populations of planets with $K\gsim10\meter\second^{-1}$. The black solid line denotes the distribution for observed exoplanets, the red dashed line for the reference population (R), the blue dot-dashed line for the population with 20 intial embryo seeds (E20) and the cyan dotted line for the population with 5 initial embryo seeds (E5).}\label{fig:comp_prat}
\end{figure}

\begin{figure}
\includegraphics[width=0.5\textwidth]{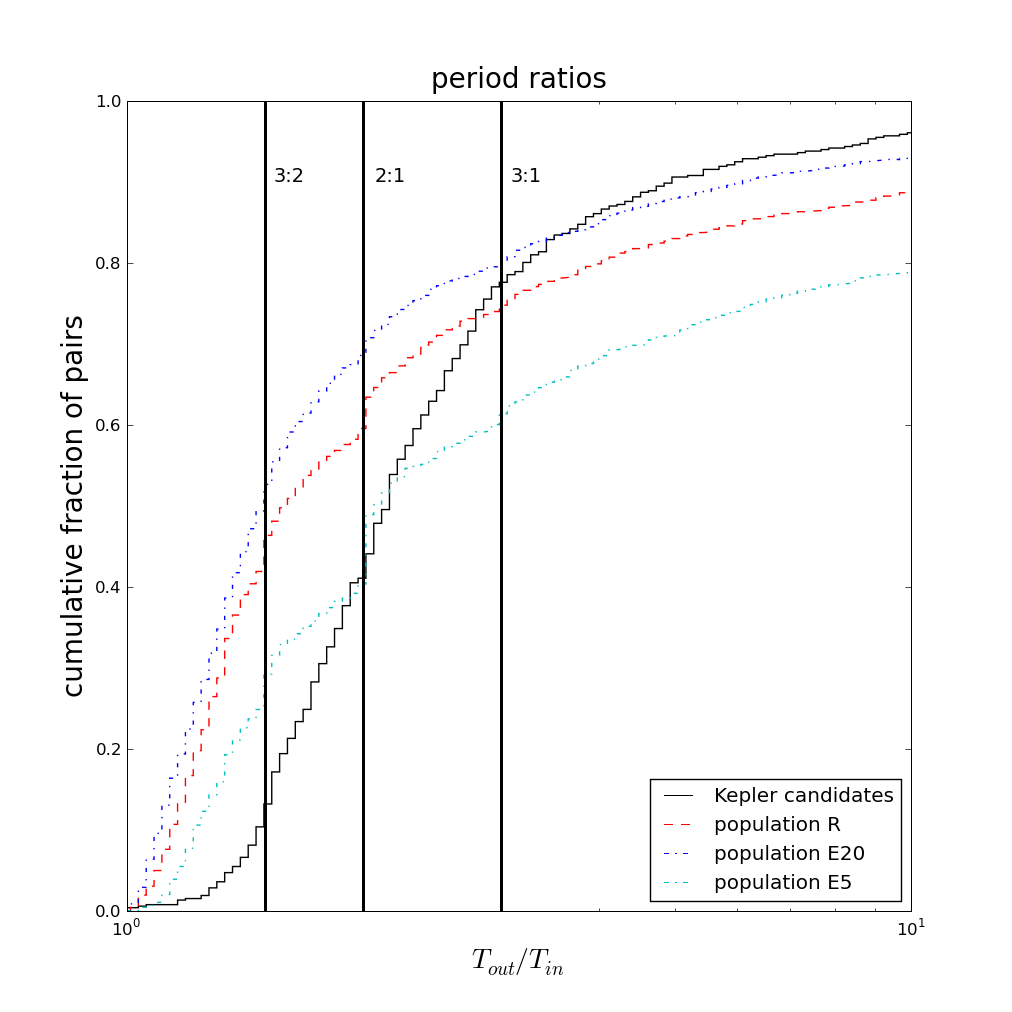}
\caption{Comparison of the period ratios after $100\Myr$ for different planet populations. The black solid line denotes the distribution for Kepler planet candidates (data taken from \textbf{www.exoplanets.org}), the red dashed line for the reference population (R), the blue dot-dashed line for the population with 20 intial embryo seeds (E20) and the cyan dotted line for the population with 5 initial embryo seeds (E5).}\label{fig:kcomp_prat}
\end{figure}

\section{Discussion}

\begin{enumerate}

\item \textit{Post-formation evolution effects -- eccentricity}\\ 
In the simulations presented in this paper, evolving various populations obtained with the Bernese planet formation code for an additional $100\Myr$ after the time of disk dispersal affects the planetary eccentricities. Populations with low initial eccentricities show an overall increase in median eccentricity resulting from increased eccentricities for low-mass planets, whereas massive planets exhibit minimal increase or even decrease in eccentricity. Contrary to the expectation that post-formation evolution leads to increased eccentricities, initially eccentric populations experience an ``eccentricity evaporation'' effect where the more eccentric planets are rapidly lost through collisions and ejections, resulting in lower median eccentricities after the post-formation evolution, but still higher than in lower initial eccentricity populations.\\
The eccentricity distributions of all populations followed for $100\Myr$ in this paper remain a poor match to observations. To further increase the final eccentricities, additional sources of excitation such as stellar fly-bys (see below) would be required in our model.

\item \textit{Post-formation evolution effects -- mass, semi-major axis and composition}\\ While the eccentricity distributions exhibit some change over the course of the post-formation evolution, the distributions of semi-major axes and planetary masses as well as the distribution of period ratios are largely unchanged by post-formation effects. The latter might be partly due to the relatively low dynamical activity in the populations we studied as there is a clear correlation between the dynamical activity (gauged by proxy of using the number of ejections and collisions and the fraction of systems exhibiting either) and the amount of change between distributions at the time of disk dispersal and after $100\Myr$ for both semi-major axis and mass distributions. This low amount of change in distributions even in the most dynamically active population for planets with $\rv$ implies that the observed distributions of exoplanet semi-major axes and masses are a close approximation to their state at the end of their respective disks' lifetimes, barring external perturbations. Similarly, the chemical compositions of massive planets and planets on wide orbits exhibit minimal changes from post-formation evolution effect as collisions (and the resulting changes in chemical composition) primarily occur between planets with $m_p\leq \Me$. Comparisons of masses and semi-major axes between planet formation results and observed exoplanets, such as M09b, therefore are a warranted approach.

\item \textit{Number of initial embryos}\\ While the initial number of embryo seeds has a minimal influence on the distribution of planetary masses and only small influence on the distribution of semi-major axes after $100\Myr$ for planets with $K\geq10\meter\second^{-1}$ in our simulations, the impact on planetary eccentricities and period ratios is very important. For period ratios, increasing the initial number of embryos decreases the fraction of planet pairs with period ratios close to the nominal location of low-order mean-motion resonances, but \textbf{this fraction} remains relatively unchanged by the post-formation evolution. 

\item \textit{Integration time}\\ 
The $100\Myr$ integration time  
is sufficiently long for dynamical instabilities to arise and resolve themselves through collisions and scatterings, and studies of dynamically active systems by e.g. \cite{Juric2008} find that dynamical activity strongly decreases after $100\Myr$. In Figure \ref{fig:tdyn}, we show the distributions of the timing of dynamical events during the post-formation evolution in the populations we studied. The median event times are $\tau=1.28\Myr$ for the reference population, $\tau=2.24\Myr$ for RD, $\tau=3.69\Myr$ for ND, $\tau=4.89\Myr$ for E5, $\tau=0.41\Myr$ for E20 and $\tau=0.39\Myr$ for RA. In all populations except for E5, where the very low number of dynamical events with $n_{event}=36$ makes statistics somewhat unreliable, $\gsim90\%$ of all events happen within the first $50\Myr$ of post-formation integration time, leading us to conclude that for most systems, $100\Myr$ represent a sufficiently long time to transition from the formation phase to long-term stable configurations.
\begin{figure}
\includegraphics[width=0.5\textwidth]{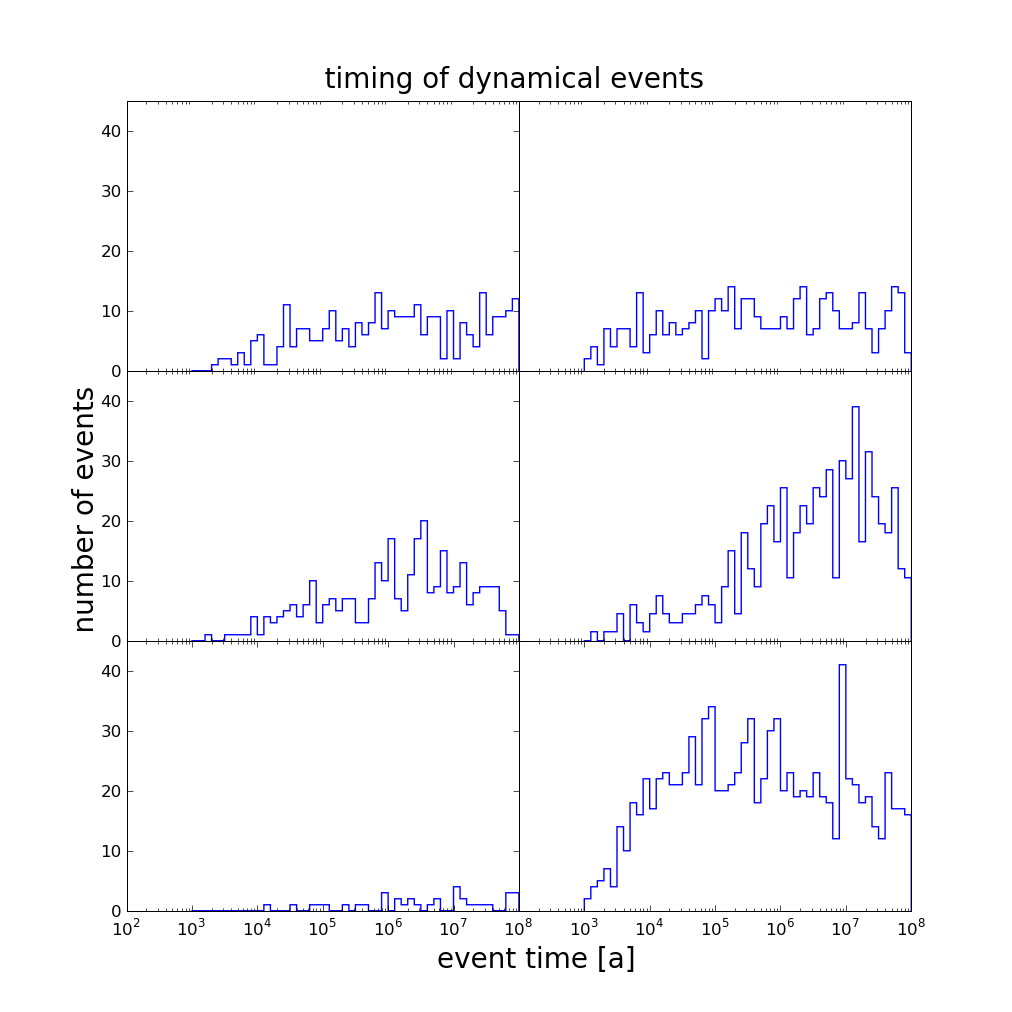}
\caption{Timing of ejections and collisions in the populations R (upper left), RA (upper right), RD (middle left), ND (middle right), E5 (lower left) and E20 (lower right).}\label{fig:tdyn}
\end{figure}

\item\textit{Free-floating planets}\\ Observations of microlensing events have resulted in claims that, in addition to exoplanets bound to their host stars, a large population of \textit{free-floating} or \textit{rogue} planets exist, see e.g. \cite{Sumi2011}. While this claim is disputed by e.g. \cite{Veras2012}, a sizeable population of planets ejected from their birth systems might still exist. In our populations, planets are ejected both during the formation phase and during the post-formation evolution. Unlike the large fraction of jovian mass planets claimed by microlensing observations, the ejected planets in our populations have median masses $m\leq2\Me$, and the fraction of ejected planets with $m\gsim\MJ$ versus the total number of planets with $m\gsim\MJ$ in a given population are $\lsim5\%$ for the populations with full eccentricity and inclination damping, $\sim19\%$ for the population with reduced damping and $\sim33\%$ for the population without damping. We therefore do not produce the large number of free-floating jovian planets expected from microlensing surveys. If a large number of free-floating planets would be confirmed, this would imply either a different formation mechanism for such planets, e.g. through gravitational instabilities in the outer disk (see \cite{Boss2011}), or a significantly larger amount of dynamical interactions during and after the formation phase.

\item\textit{Comparison with previous studies}\\ The low number of ejected giant planets is at odds with assumptions on the initial conditions in studies reproducing the observed exoplanet eccentricity distributions from giant planet scattering, such as e.g. \cite{Juric2008}, \cite{Chatterjee2008} or \cite{Malmberg2009} (referred to as JT08, C08 and MD09, respectively). In all three studies, the initial planetary systems are more massive, with multiple planets with masses between $0.1\MJ$ and $10\MJ$, and more compactly spaced, with median separations in mutual Hill radii $b=\frac{a_{O}-a_I}{R_{H,IO}}$, where $R_{H,IO}=\frac{a_I+a_O}{2}\sqrt[3]{\frac{m_I+m_O}{3M_{\star}}}$, ranging from $b = 2$ in MD09 to $b=4.4$ in C08 and active systems in JT08. In contrast, the planets with $\rv$ in our reference population have a (slightly) larger median separation of $b=5.2$. The timescale for orbit crossing for a given planet pair strongly depends on the orbital separation (see e.g. \cite{Zhou2007}), such that even the moderate increase in separation between planets results in a strongly reduced amount of dynamical activity in our simulations. Moreover, the assumptions on planetary masses, especially on the frequent occurrence of multiple jovian-mass planets in a single system, are not reproduced by any population in our simulations. Indeed, the fraction of systems with more than one planet with $m_p\geq30\Me\simeq 0.1\MJ$ in our reference population is $\sim22\%$, the fraction of systems with more than two such planets is only $\sim3\%$. If we set the mass limit to $m_p\geq 1\MJ$, the respective fractions drop to $11\%$ and $1\%$. Clearly, the assumptions on the initial conditions in C08, JT08 and MD09 are not reproduced by the  planet formation model we use. \\
As in \cite{Thommes2008}, we do find an initial population of eccentric giant planets from the formation process. The populations RD and ND (with reduced and no eccentricity damping) in particular produce several eccentric giant planets and show the best initial agreement with observed exoplanet eccentricities. For both populations, however, the final eccentricities are reduced by the efficient removal of eccentric planets by means of ejections and collisions, resulting in less agreement with observed exoplanet distributions.

\item \textit{External perturbation as a source of dynamical activity}\\ With the exception of the population without eccentricity and inclination damping, our populations have typical eccentricities much lower than those of observed exoplanets. Despite uncertainties on whether some of the observed highly eccentric exoplanets could be better explained by multiplanet systems with smaller eccentricities, see e.g. \cite{Angela-Escude2010} or \cite{Wittenmyer2013}, the post-formation evolution of our populations is insufficient to reproduce observed eccentricities, thus requiring additional sources of eccentricity. In our simulations, we consider the forming systems to be isolated, whereas star formation and, by consequence, planet formation typically takes place in clusters. Studies by e.g. \cite{Malmberg2007b} and \cite{Malmberg2011} have shown that in their reference cluster, a solar-mass star on average has four close encounters\footnote{A close encounter in this context is defined as an encounter with $r_{min}\leq1000\AU$, where $r_{min}$ is the minimum separation of the two stars.} and $\sim75\%$ have at least two close encounters. Roughly half of these close encounters take place during the first $10\Myr$, which is of the same order as the disk lifetimes (see \cite{Mamajek2009}), and $\sim90\%$ happen within $100\Myr$. These stellar fly-bys can affect the dynamical evolution of nascent planetary systems in two ways: by either directly scattering or ejecting planets in the case of very strong encounters, or by reducing the stability of the system by increasing eccentricities and/or inclinations and potentially force planets out of stabilizing mean-motion resonances. \cite{Malmberg2011} found that between $\sim40\%$ and $\sim60\%$ of their systems underwent ejections or orbit-crossings after a stellar fly-by, implying that the contributions from stellar encounters to the final architecture of planetary systems could be significant. The inclusion of stellar fly-bys in the current planet formation model will be the topic of future studies.

\end{enumerate}

\section{Conclusion}

Within the first $100\Myr$ of evolution after the dissipation of the protoplanetary disk, planetary systems can undergo dramatic changes in orbital architecture through planet-planet scattering, ejections and collisions. These dynamical interactions are necessary to bring the eccentricities of planet populations in the Bernese planet formation model in line with observed exoplanet eccentricities. However, the effect of the post-formation evolution on the eccentricity distributions alone remains insufficient to explain observed eccentricities, requiring further investigation into the assumptions on initial conditions as well as the inclusion of additional sources of dynamical perturbation such as stellar fly-bys in our model.\\
While the eccentricities clearly show the effects of post-formation evolution, other planetary characteristics such as the mass and semi-major axis distributions remain largely the same after $100\Myr$ as at the time of disk dispersal, implying that the mass and semi-major axis of a planet are by and large determined by the formation process alone.

\begin{acknowledgements}This work has been in part carried out within the frame of the National Centre for Competence in Research PlanetS 
supported by the Swiss National Science Foundation. The authors acknowledge the financial support of the SNSF. This work was supported by the European Research Council under grant 239605.\end{acknowledgements}

\bibliographystyle{aa}

\addcontentsline{toc}{chapter}{Bibliography}
\bibliography{phdbib}

\end{document}